\documentclass[preprint,10pt]{aastex}

\slugcomment{Accepted to ApJ} 

\shorttitle{The Her X-1 Disk: X-ray Spectroscopy and Modeling}
\shortauthors{Jimenez-Garate et al.}
\def\chandra{{\it Chandra }}
\def\xmm{{\it XMM-Newton }}
\def\xmms{{\it XMM-Newton}}

\def\hst{{\it HST }}

\def\degr{$^{\circ}$}

\def\cm{{\rm\thinspace cm}}

\def\erg{{\rm\thinspace erg}}

\def\keV{{\rm\thinspace keV}}
\def\km{{\rm\thinspace km}}

\def\Msun{\hbox{$\rm\thinspace M_{\odot}$}}

\def\ph{{\rm\thinspace ph}}
\def\s{{\rm\thinspace s}}

\def\pcmcu{\hbox{$\cm^{-3}\,$}}

\def\ergpcmsqps{\hbox{$\erg\cm^{-2}\s^{-1}\,$}}

\def\ergps{\hbox{$\erg\s^{-1}\,$}}

\def\kmps{\hbox{$\km\s^{-1}\,$}}

\def\pcmsq{\hbox{$\cm^{-2}\,$}}

\def\phpcmsqps{\hbox{$\ph~\cm^{-2}\s^{-1}\,$}}

\def\powerlawfluxat1kev{\hbox{$\ph~\cm^{-2}\s^{-1}\keV^{-1}$}}

\def\nH{$N_{\rm H}\,$}

\begin{document}


\title{Identification of an Extended Accretion Disk Corona in the Hercules X-1 Low State: 
Moderate Optical Depth, Precise Density Determination, and Verification of CNO Abundances} 


\author{M. A. Jimenez-Garate}
\affil{MIT Center for Space Research, 70 Vassar St (NE80-6009), Cambridge, MA 02139}
\email{mario@space.mit.edu}

\author{J. C. Raymond}
\affil{Center for Astrophysics, 60 Garden St., Cambridge, MA 02138}
\email{raymond@cfa.harvard.edu}

\author{D. A. Liedahl}
\affil{Lawrence Livermore National Laboratory, 
Department of Physics and Advanced Technologies, 7000 East Ave., L-41, Livermore, CA 94550}
\email{liedahl1@llnl.gov}

\and

\author{C. J. Hailey} 
\affil{Columbia Astrophysics Laboratory, 538 W. 120th St., New York, NY, 10027}
\email{chuckh@astro.columbia.edu}


\begin{abstract}
We identify an accretion disk atmosphere and corona
from the high resolution X-ray spectrum of Hercules X-1, and we
determine its detailed physical properties.
More than two dozen recombination emission 
lines (from \ion{Fe}{26} at 1.78~\AA \ to \ion{N}{6} at 29.08~\AA) 
and Fe K$\alpha$,K$\beta$ fluorescence lines were detected in a 50~ks observation with
the \chandra High-Energy Transmission Grating Spectrometer (HETGS). 
They allow us to measure the density ($n_e = 2 \pm 1 \times 10^{13}$~cm from \ion{Mg}{11}), 
temperature ($kT=7 \pm 3$~eV from \ion{Ne}{9}), spatial distribution, 
elemental composition, and kinematics ($\Delta v \lesssim 260$\kmps) of the plasma. 
We exclude HZ~Her as the source of the recombination emission.
We compare accretion disk model atmospheres with the observed spectrum in 
order to constrain the stratification of density and ionization, 
elemental composition, energetics, and thermal stability.
The derived disk atmosphere and corona radii are
$8 \times 10^{10} \lesssim r \lesssim 1 \times 10^{11}$~cm, in agreement with 
previously measured eclipse ingress light curves. 
The atmospheric spectrum observed during the low state 
is photoionized by the main-on X-ray continuum, 
indicating that the disk is observed edge-on during the low state.
We infer the mean number of scatterings $\langle N \rangle$ of Ly$\alpha$ and Ly$\beta$
line photons from H-like ions. We derive $\langle N \rangle \lesssim 69$ for 
\ion{O}{8} Ly$\alpha_1$, which rules out the presence of a mechanism 
modeled by Sako (2003) to enhance 
\ion{N}{7} emission via a line overlap with \ion{O}{8}.
The line optical depth diagnostics are consistent with a flattened atmosphere. 
Our spectral analysis, the disk atmosphere model,
and the presence of intense \ion{N}{7} and \ion{N}{6} lines (plus \ion{N}{5} in the UV),
confirm the over-abundance of nitrogen relative to other metals, 
which was shown to be indicative of CNO cycle processing in a massive progenitor. 
The spectral signatures of a thermal instability in the photoionized
plasma are not evident, but the measured density is in 
the stable regime of the models. 

\end{abstract}




\section{Introduction}
High-resolution spectroscopic observations of X-ray binaries
have shown in some cases the presence of extended X-ray emitting plasmas or
outflows surrounding the accretion disk. The detection of these components
and their kinematic properties has been made possible by the observation of
emission or absorption lines from very highly ionized metals. 
In the case of Hercules X-1, 
a bright intermediate-mass X-ray binary 
with $P_{\rm orb} = 1.7$~day orbital period and
a $P_{\rm pulse} =1.24$~s X-ray pulsar \cite[]{discovery_pulsar,batse},
observations with the \xmm Reflection
Grating Spectrometer (RGS) showed a strong
35~d phase dependence in the high-resolution spectrum
\cite[]{herx1me}. The $P_{\Psi} \sim 35$~d pseudo-periodic cycle \cite[]{dipsfound} has been 
associated with a tilted and
precessing accretion disk \cite[]{precessing_disk_hyp,twisted_disk_flaps,disk_model}.
Through the 35~d cycle, the X-ray light-curve is asymmetric and contains two maxima:
a state of $\sim 8$~d duration reaching the peak flux $F_{\rm max}$
named the {\it main-on},
and a secondary high state of $\sim 4$~d duration reaching $\sim 1/3~F_{\rm max}$ 
named the {\it short-on}. A {\it low state} with $\sim 1/20~F_{\rm max}$ 
ensues at other epochs. The main-on spectrum exhibits a strong continuum, with evidence for weak
and very broad emission lines, while the low state
shows bright and narrow emission lines dominating over a very weak continuum.
The short-on is intermediate, with a moderate continuum flux and 
more intense emission lines than the low state \cite[]{herx1me}. 
The Fe K$\alpha$ line follows a similar pattern. {\it XMM-Newton} EPIC data show a 6.4~keV Fe K$\alpha$ line 
which is practically unresolved during the low and short-on states, and a broad line at 
6.5~keV with $330 \pm 20$~eV FWHM during the main-on \cite[]{ramsay}.
Matter close to the magnetosphere and the pulsar appears to be observable during the main-on only.
This behavior supports the picture of a precessing accretion disk, since
the spectroscopic appearance of the disk atmosphere and the pulsar should be highly 
dependent on inclination.
Analysis of the RGS spectrum showed a clear over-abundance of N with respect to Ne, as well
as a moderate under-abundance of O and C with respect to Ne. 
The N over-abundance was proposed as a signature of CNO cycle nucleosynthesis occurring
at an early phase in the binary system, and it is indicative of a
star twice or more times more massive than the secondary/companion \cite[]{herx1me}.
This is indicative of a massive progenitor of the system and a period of
extreme mass loss, as had been predicted by evolutionary models of LMXBs \cite[]{rappaport}.
\cite{sakorad} suggested that the N over-abundance is not real, but that it 
is due instead to an X-ray Bowen fluorescence effect which occurs at large 
line optical depths.  In this article, we quantify the relevant line optical 
depths to distinguish between the X-ray Bowen effect or the abundance anomaly scenarios. 

Hercules X-1 has been observed extensively. 
Optical light-curves \cite[]{optical_curves}
and X-ray eclipses \cite[]{discovery_pulsar} yield
a $P_{\rm orb} = 1.7$~day orbital period.
The $1.5 \pm 0.3~M_{\sun}$ neutron star has a $2.3 \pm 0.3~M_{\odot}$ companion, HZ~Her, 
which changes from A to B spectral-type over the orbital period, due to the strong X-ray
illumination on its surface \cite[]{reynolds}.
The unabsorbed luminosity of Her X-1 is $L=3.8 \times
10^{37}$~ergs$^{-1}$, using a distance of $D = 6.6 \pm 0.4$~kpc 
\cite[]{reynolds}. The Her X-1 broadband X-ray spectrum during the main-on state
consists of a blackbody component
with temperature $kT \sim 90$~eV, plus a power-law
component with a 24~keV exponential cutoff, and a
42~keV cyclotron feature \cite[]{beppo_cyc}. 
The X-ray light-curve, the variations in the pulse profiles \cite[]{gingapulse}, and the 
variability of the dips \cite[]{dipsfound,xteobs}, are fit by a geometric model
of a precessing, warped accretion disk, with an $85^{\circ}$ inclination with respect to the line of sight, 
a $20^{\circ}$ precession opening angle for the outermost disk,
and an $11^{\circ}$ precession angle for the innermost disk
\cite[]{disk_model}. 
The ultraviolet (UV) spectrum exhibits line emission from 
\ion{C}{5}, \ion{N}{5}, and \ion{O}{5}, which have two separate velocity components.
A narrow UV line component is thought to originate on the illuminated face of HZ~Her.
The broad UV line region likely originates in a prograde accretion
disk of $\sim 10^{11}$ cm radius \cite[]{uv_disk_lines}.
One of the questions to be resolved is whether the X-ray line emission
has a similar origin to the UV. The broadening of the X-ray line emission indicates
that it can originate at radii similar to the UV. For the case
of an irradiated disk atmosphere, the UV line region is denser and deeper 
than the X-ray line region.

In this work, we interpret the broadband (1.5~\AA \ to 30~\AA) high-resolution X-ray
spectrum of Her X-1 and investigate the nature of the line
emitting region(s). The \chandra grating spectrum allows us to detect a wealth of
spectral features which had not been previously observed in Her X-1.
Our data reduction procedure is described in \S \ref{sec:obs}.
To interpret the spectrum, we use two complementary approaches: 1) an analytic 
approach for constraining the physical parameters of the plasma, based solely on 
the photoionized plasma physics; and 
2) a synthetic approach, by comparing the spectrum with that of an astrophysical model. 
In \S \ref{sec:analysis}, discuss the origin of the Fe K$\alpha$ line and
we perform spectral diagnostics on density, kinematics, and location from the recombination
lines.
In \S \ref{sec:em}, we utilize standard photoionized plasma models and 
radiative recombination rates to perform an emission measure analysis,
and to quantify the ionization distribution of the plasma, the emitting volume, 
and the ratios among the metal abundances.  We also compare the model
temperatures with those measured from radiative recombination continua (RRC).
In \S \ref{sec:odepth}, we quantify the mean number of scatterings of selected
lines, which shows that the optical depth is 
not high enough for the X-ray Bowen fluorescence effect to take place in Her X-1. 
Line depth diagnostics also provide evidence for an anisotropic plasma distribution.
In \S \ref{sec:modcomp}, we identify the nature of the X-ray emission region 
on the basis of a comparison of an accretion disk and corona model with the observed
spectrum. In \S \ref{sec:stab}, we discuss the thermal stability of the plasma.  Finally, in 
\S \ref{sec:adc} we compare the Her X-1 spectrum with that of 
Accretion Disk Corona (ADC) sources, and we conclude in \S \ref{sec:concl}.

\section{\chandra Observations}
\label{sec:obs}
Hercules X-1 was observed on 2002 May 5 at 10:15 UT
with the \chandra High Energy Transmission Grating 
Spectrometer \cite[HETGS]{hetg}.
The observation lasted 49.4~ks. Data from the
All-Sky Monitor (ASM) instrument onboard the
Rossi X-ray Timing Explorer (RXTE), 
indicated that Her X-1 was in the low state
($\Psi = 0.44-0.46$) while at mid-orbit
($\phi = 0.33-0.67$)
during the \chandra observation.
From the RXTE ASM data, we determined that
first main-on start signal immediately preceding
the Chandra observation occurred on
MJD 52384.22. \cite{xteobs} observed that 
main-on turn-ons occur only orbital phases
0.23 or 0.68, possibly due to a disk-orbit
locking mechanism. The closest, most likely
main-on start time occurred at orbital phase 0.23,
at MJD 52383.956, $\sim 10$~hr earlier than 
the RXTE ASM detection.

We processed the HETGS data with the
CIAO analysis software version 3.0.
We obtained Medium Energy Grating (MEG) and High Energy
Grating (HEG) spectra, which have 
energy resolution FWHM of $\Delta \lambda = 0.023$~\AA \ and
$\Delta \lambda = 0.012$~\AA, respectively.
The ACIS-S CCD was set in FAINT mode since the low state does not
produce pile-up in the image, and this minimizes the background 
in the spectrum. 
We use the {\it destreak} tool to clean the ACIS-S
image. The {\it mkrmf} and {\it mkarf} tools 
are used to produce the response matrices and effective area
applicable to our observation. 
We use the calibration data CALDB version 2.26,
which was updated 2004 Feb 2 to correct for the contamination layer
on ACIS-S. We used the ISIS spectral analysis software in our model fitting and flux
measurements \cite[]{isis}. The {\it lightcurve} routine
was used to extract the time variability of the flux.
Her X-1 was placed at a Y Offset of 0.33~arcmin, 
which is practically on-axis.

\section{Spectral Analysis}
\label{sec:analysis}

\begin{figure}
\epsscale{0.7}
\plotone{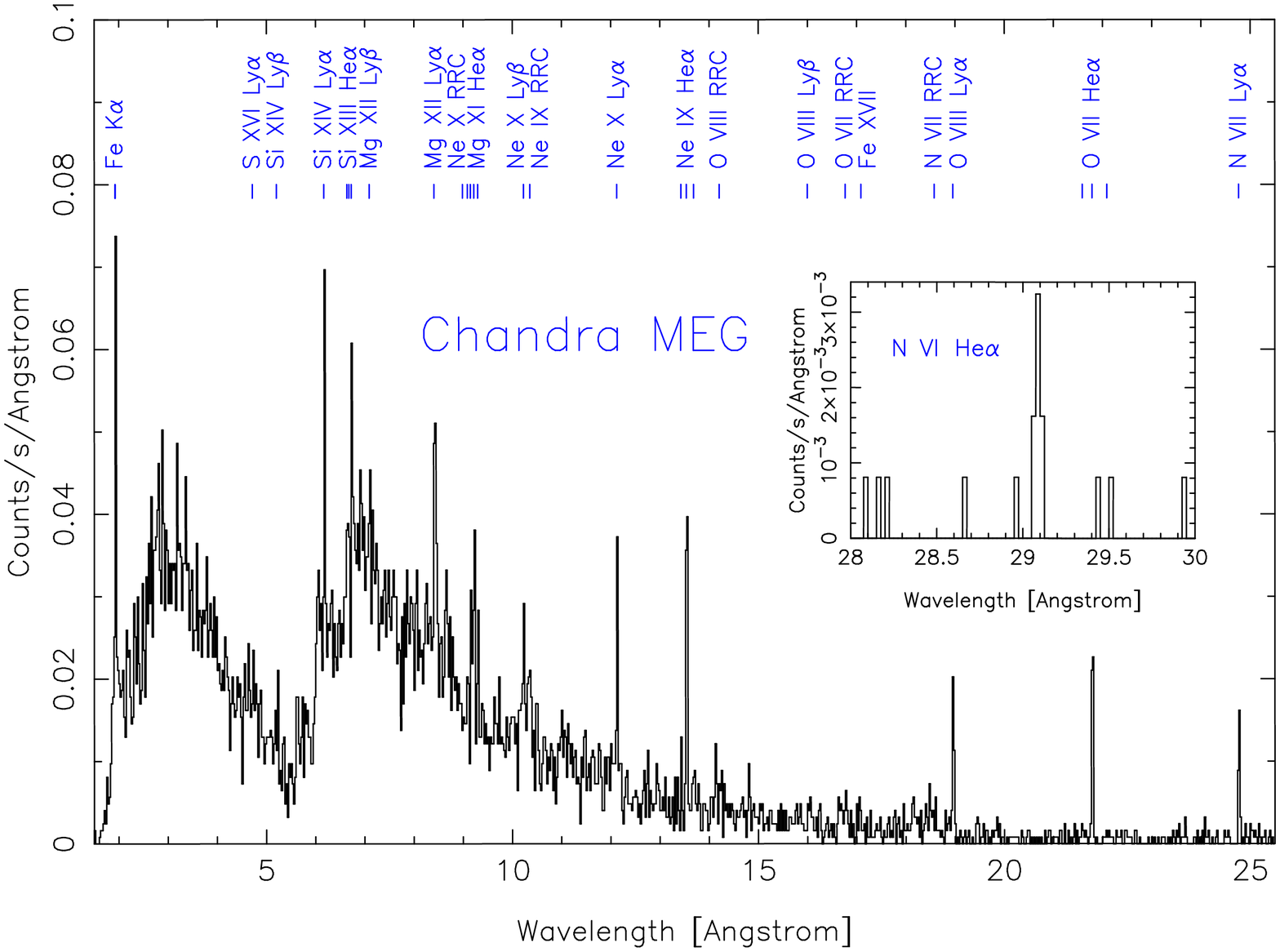}
\plotone{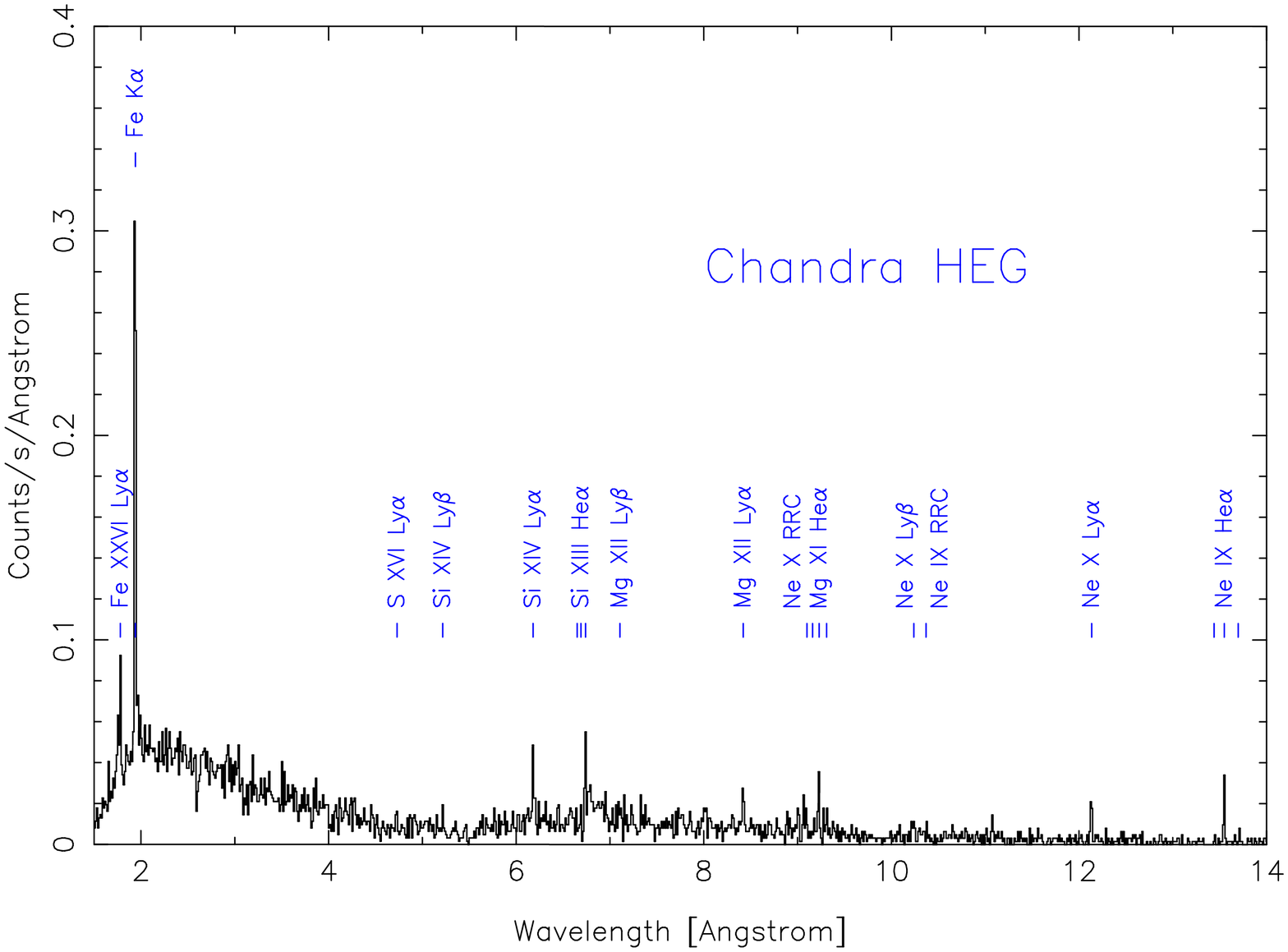}
\caption{The MEG and HEG count spectra of Hercules X-1 during its low-state
shows numerous radiative recombination lines and continua, a strong
Fe K$\alpha$ fluorescence line, and a continuum which may be produced by Compton scattering.
The large range of temperatures at which these emission lines originate
is a direct probe of the ionization structure of the upper layers of the accretion
disk atmosphere and corona.
\label{fig:megtot}}
\end{figure}

The HEG and MEG spectra of Her X-1 exhibit emission lines throughout the X-ray band. 
Together, the spectra exhibit bright lines from the H-like and He-like ions of
N, O, Ne, Mg, and Si, plus the lines from H-like ions of S and Fe. 
The MEG and HEG counts spectra in Figure \ref{fig:megtot} show a 
strong Fe K$\alpha$ fluorescence 
line, as well as numerous recombination emission lines.
The HEG spectrum in Figure \ref{fig:fe} highlights the 
Fe K$\alpha$ line from low ionization states
and a line from the high ionization state \ion{Fe}{26}.
The spectra of HEG and MEG are added for display purposes in Figure
\ref{fig:megheg}, and the MEG spectrum at the high wavelength end is shown
in Figure \ref{fig:megon}. 
The detection of RRCs, as well as the 
fact that weak Fe L lines 
are observed, are characteristic of photoionized plasmas \cite[]{liedahl,liedahl99}.
We measure a 0.5-7~keV flux of $F_{0.5-7} = 3.3 \times 10^{-11}$\ergpcmsqps,
and the fluxes of individual spectral features are listed in Table \ref{tab:lines}.
We use Cash statistics to fit the individual spectral features and obtain the
statistical errors. Continuum fit parameters are shown in Table \ref{tab:cont}.

\subsection{Fe fluorescence}

The data allow us to fit independently the K$\alpha_1$ and K$\alpha_2$ lines, which are
merged in a single feature. We also detect the K$\beta$ line.
The measured Fe K fluorescence line energies, as well as the line ratios 
among K$\alpha_1$, K$\alpha_2$, and K$\beta$, all show that the iron is neutral,
based on the \cite{meka} calculations.  The K$\alpha$/K$\beta$ ratio is $6.2 \pm 1.6$ (compared to 7.99
for \ion{Fe}{1} from Kaastra \& Mewe), while
the K$\alpha_1$/K$\alpha_2$ ratio is $2.28 \pm 0.45$ (compared
to 2.00, idem). 
Any iron ion with an L-shell electron 
can fluoresce with high yields as well. However, 
the line energy observed corresponds to \ion{Fe}{1}--\ion{Fe}{9}
\cite[]{fe1to9}, and likely up to \ion{Fe}{13}.
We approximate the 6.4~keV fluorescence line flux (in \phpcmsqps) with
\begin{equation}
F \thickapprox \onehalf ~Y_{\rm Fe}~f~T ~ \Bigl( \frac{\Omega}{4\pi} \Bigr)  \int_{7.1~{\rm keV}}^{200~{\rm keV}} F_E ~(1 - e^{-\tau_{\rm Fe}}) ~dE
\end{equation}
where we take $F_E$ to be the continuum flux (in \powerlawfluxat1kev) observed during the main-on by \citet{beppo_cyc},
$\tau_{\rm Fe}$ is the optical depth of M-shell iron, $Y_{\rm Fe} = 0.34$ is the fluorescence
yield of Fe, $\Omega$ is the solid angle subtended by the fluorescing plasma
from the vantage point of the X-ray pulsar, 
$0<T<1$ is a transmission coefficient due to absorption 
from species other than \ion{Fe}{1}--\ion{Fe}{13} and Compton scattering,
and $f$ is the fraction of the fluorescing region which is visible to us.
Roughly half the fluorescent photons are emitted downward into the optically thick gas.
The 0.1 to 100 keV continuum flux during the main-on is 
$\int F_E dE = 6.7 \times 10^{-9}$\ergpcmsqps, and it is
taken from the Model 2 fit by \citet{beppo_cyc}, which consists of an absorbed 
blackbody plus broken power law, with an exponential cutoff and cyclotron absorption line. 
Using $\tau_{\rm Fe} \sim 4$, 
which is an estimate of the depth from which an Fe K photon can escape,
we obtain $f T \Omega \sim 0.22$.
The solid angle subtended by the companion star is $\Omega \sim 0.50$,
and the solid angle subtended by the disk photosphere is
$\Omega \sim 1.0$, as obtained from the illuminated disk model \cite[]{diskmodel}.
These $\Omega$ are quite similar. 
Both cases produce $f T \sim 0.2$--0.4. Given $0.33 < \phi < 0.67$ during
the observations, we estimate that $f \sim 2/3$ of the illuminated star is visible. For the disk,
$f \lesssim 1/2$ for the scenario of a self-shielding
edge-on disk (see next section). In both cases 
reasonable transmission $T \gtrsim 0.5$ can be obtained.
Thus, we need to employ the orbital variability to distinguish amongst these components (\S \ref{sec:var}).
The Fe K$\alpha$ fluorescence line flux and broadening are much larger during
the main-on than during the short-on, and the low state line flux measured
with \xmms, of (2--$12) \times 10^{-4}$ \phpcmsqps \cite[]{zane}, brackets the Fe K$\alpha$ line flux we measure with HETGS (see Table \ref{tab:lines}).

\subsection{Unbroadened Lines} 
\label{sec:broad}

The gratings do not resolve any Doppler width in the emission lines. 
The Doppler broadening $(\Delta v)_\sigma$ is measured by fitting Gaussian 
profiles to the brightest lines 
in the spectra, such that $(\Delta v)_{\rm FWHM} \simeq 2.35 (\Delta v)_\sigma$. Table \ref{tab:lines} shows that the best upper limits on 
$\Delta v$ are placed on the lines
with largest $\lambda$. The observed line broadening is entirely attributed to their doublet
nature.
We set 90 \% confidence limits of $(\Delta \lambda)_\sigma =0$--8~m\AA \ for the
\ion{O}{7} $i$ line with MEG, $(\Delta \lambda)_\sigma = 8$--17~m\AA \ for the 
\ion{O}{8} Ly$\alpha$ line with MEG, 
and $(\Delta \lambda)_\sigma = 0$--3~m\AA \ for the \ion{Ne}{9} $i$ line with HEG. 
The remaining bright lines have similar $(\Delta \lambda)_\sigma$.  The intrinsic line
broadening, plus any residual instrumental broadening for MEG and HEG not
accounted for in the response matrix, is quantified using the 
spectra of a bright star from which no velocity broadening is expected. 
We choose the star HR~1099 as reference, which yields 
$(\Delta \lambda)_\sigma = 4.8 \pm 0.4$~m\AA \ 
with HEG and $(\Delta \lambda)_\sigma = 5.2 \pm 0.6$~m\AA \ with MEG
at 12.13~\AA \ (99\% confidence limits). By fitting a plasma model, we
verify that this broadening can be fully accounted for by the 
\ion{Ne}{10} Ly$\alpha$-doublet, which is separated by $5$~m\AA \
and should show a 2:1 intensity ratio.  The doublet fully accounts for the measured $(\Delta \lambda)_\sigma$.
Other lines in the HR~1099 spectrum are consistent with this.
The upper limits on the velocity 
broadening on Table \ref{tab:lines} are based on either
the measured or reference $(\Delta \lambda)_\sigma$, whichever is largest.
At the slightly lower resolving power of the \xmm RGS, 
the short-on and low state emission lines
of Her X-1 were unresolved \cite[]{herx1me}.  

The absence of a Doppler velocity broadening constrains the dynamics of
the disk atmosphere and the geometry of the disk. Since the disk inclination
is $i \sim 85$\degr, the orbital velocity of the disk is given by 
$v \sim \sqrt{G M / R} ~ \cos \Phi$, 
where $G$ is the gravitational constant, $M$ the neutron star mass, $R$ the 
disk radius, and $\Phi$ is the azimuth on the disk.
If the lines produced by a $\Phi$-symmetric disk,
translating the measured $(\Delta v)_\sigma$ into a physical size of the 
disk requires modeling the line profiles with a calculated (or assumed)
emissivity versus radius.  The disk lines would have double-peaked profiles.
The profiles from a centrally illuminated disk of 
$10^{8.5} < r < 10^{11}$~cm radius in full view calculated with the \citet{diskmodel} model 
can be fit with a $(\Delta v)_\sigma \sim 750$~\kmps Gaussian.
Clearly the disk cannot be in full view in the Her X-1 low state. 
{\it This is an indication that we are observing the outer rim of the disk atmosphere 
and corona}.
For a $\Phi$-symmetric, edge-on, and optically thin disk,
the emission region radius would be 
$r \gtrsim G M / [ 4 (\Delta v)_\sigma^2 ]$.
For the X-ray lines observed at highest resolving power,
the inferred lower limits with this equation are
$r \gtrsim (2$--$3) \times 10^{11}$~cm, 
larger than the neutron star Roche lobe 
and the disk size $r \sim 1.4 \times 10^{11}$~cm deduced from the eclipse
ingress light curves of the
\ion{He}{2}, \ion{C}{4}, \ion{Si}{4}, and \ion{O}{5} emission lines in the UV
\cite[]{uv_lines_wind_model,uv_disk_lines}. Therefore, 
the disk atmosphere is likely asymmetric.
One interpretation is that due to the flared disk geometry, the 
disk shields itself, such
that we only observe the far side of the disk, which causes the $\cos \Phi$ factor 
to reduce $\Delta v$. 
This does not require an asymmetric disk, just one that partially shields
itself. Another interpretation is that
the disk is asymmetric or warped, which can similarly reduce 
the observed $\Delta v$. Evidence of shielding in a warped disk
was found by \cite{uv_lines_wind_model} in the UV emission line profiles.
A third possibility is that the disk is
$\Phi$-symmetric, but has an atmosphere which is orbiting at sub-Keplerian velocities.
The case of the asymmetric disk implies a modulation of the X-ray line fluxes 
with orbital phase (see \S \ref{sec:var}).

\subsection{A location and density diagnostic from He-like ions}
\label{sec:helike}

The He$\alpha$ line triplets are shown in Fig. \ref{fig:triplet}.  At the HETGS resolution, each He$\alpha$ 
complex consists of the intercombination ($i$), forbidden ($f$), and resonance ($r$) lines.
The He-like triplet diagnostics were performed 
with the \xmm RGS spectrum for \ion{N}{6}, \ion{O}{7}, and \ion{Ne}{9} \cite[]{herx1me}.
Both RGS and HETGS spectra show that the $R = f/i$ flux ratio is zero for the latter ions.
The HETGS spectrum reveals that the \ion{Mg}{11} and \ion{Si}{13} He$\alpha$
lines behave differently: the $R$ ratio is nonzero and increasing with $Z$ (see Table \ref{tab:helike}). 
The value of the $R$ ratio depends on the atomic kinetics in each ion.
The $f$ line can get converted into the $i$ line by a process in which
the metastable $1s2s\ ^3S_1$ level is excited to $1s2p\ ^3P_{1,2}$ \cite[]{porquet}.
This occurs when the excitation rate $w_{\rm f \to i}$ is larger than the decay rate $w_{\rm f}$
of $1s2s\ ^3S_1$ to the ground state. We note that the particular $1s2s\ ^3S_1 \to 1s2p\ ^3P_{2}$ 
transition does not 
modify the $R$ ratio for the low and mid-$Z$ ions, for which
$1s2p\ ^3P_{2}$ decays back quickly to 
$1s2s\ ^3S_1$, instead of decaying to the ground state.
An $R \sim 0$ ratio can be produced by either collisional excitation
above a critical electron density $n_e^{\rm crit}$ \cite[]{porquet,bauti} or by photoexcitation 
above a critical UV-photon flux $F_\nu^{\rm crit}$ \cite[]{mewe}. 
We discuss both as limiting cases for which $R \to 0$.
If both the plasma density and the ambient UV flux are below their 
thresholds, the $R$ ratio reaches an asymptotic value of $R \gtrsim 2.2$, which increases with $T$ 
and depends weakly on $Z$. For \ion{Mg}{11} at $T = 3 \times 10^5$~K, the maximum 
$R = 2.8$ \cite[]{porquet}. 

We first discuss the limiting case in which the density is below threshold
and the UV flux is above threshold.
From the measured UV flux $F_\nu$, we constrain the distance ($d$) between the 
X-ray line emission region and the UV source. 
The dominant source of UV flux is the surface of the companion (see below).
A similar diagnostic was used by \citet{kahn}. 
\citet{herx1me} set a limit of $d < 7 \times 10^{11}$~cm for Her X-1.
With the newly detected lines, $d$ is shown to be equal or larger than the binary separation. 
To measure $d$, we quantify the photoexcitation rate by using the UV fluxes that were 
observed with the Hubble Space Telescope (\hst) an the Far Ultraviolet Spectroscopic Explorer or FUSE \cite[]{ghrs, uv_disk_lines, fuse}, and with
the Hopkins Ultraviolet Telescope \cite[]{hut_uvlines}. 
The relevant photoexcitation rate is: 
\begin{equation}
\label{eq:pe}
w_{\rm f \to i} = \frac{\pi e^2}{ m_e c} F_{\nu_{f \to i}} f_{\rm osc},
\end{equation}
where $e$, $m_e$ are the electron charge and mass, $c$ is the speed of light,
$f_{\rm osc}$ is the oscillator strength, and
$F_{\nu_{f \to i}}$ is the flux (in photons~s$^{-1}$~cm$^{-2}$~Hz$^{-1}$)
at frequencies resonant with the $1s2s\ ^3S_1 \to 1s2p\ ^3P_{1}$ transition in
the UV. We use the $f_{\rm osc}$ calculated with the
HULLAC atomic code \cite[]{hullac}, and those calculated by \citet{cann}.
Since $F_{\nu_{f \to i}}$ is a function of $d$, we constrain 
$d$ by relating $w_{\rm f \to i}$ to $w_{\rm f}$.

In the second limiting case, the UV flux is below threshold and the density is above threshold.
In this case, the lower limit of $n_e$ can be obtained from the upper limit on $R$ using
the  \cite{porquet} calculations.
The $d$ calculated with Eqs. \ref{eq:pe} and \ref{eq:bal}
in the first limiting case, as well as the inferred $n_e$ in the second limiting case,
are shown in Table \ref{tab:helike}.

A valuable new constraint is inferred from the \ion{Mg}{11} He$\alpha$ lines, which show
a critical $R$ ratio, that is, the value of $R = 0.52 \pm 0.13$ is intermediate between the 
asymptotic values of $R=0.0$ and $R=2.8$.
The measured $R$ ratio implies that 
\begin{equation}
\label{eq:bal}
w_{\rm f \to i} \sim w_{\rm f}.
\end{equation}
For He-like ions other than \ion{Mg}{11}, Eq. \ref{eq:bal} becomes an inequality. 
We interpret the \ion{Mg}{11} $R$ ratio in two limits and in the general case.
In the low density limit, 
the UV flux is at a critical value, and
$F_\nu$ is measured.  In the low $F_\nu$ limit, the density is at a critical
value, and the 
measurement of $n_e$ is applicable.  In the general case, the same numbers for \ion{Mg}{11} 
become a lower limit for $d$ and an upper limit for $n_e$. 
A confidence region in ($d$,$n_e$)
space requires new atomic kinetics calculations.
Since in the general case $d \ge 2 \times 10^{11}$~cm, we conclude that {\it the X-ray 
recombination line emission
cannot possibly originate in the illuminated atmosphere of HZ~Her.} 
This does not preclude an Fe~K$\alpha$ {\it fluorescence} line to be 
produced on HZ~Her.
If the recombination lines did originate within the strong UV field of the companion,
the \ion{Mg}{11} and \ion{Si}{13} would have shown no $f$ line and a strong $i$ line,
like the rest of the He-like ions.  
The caveat for the $d$-limits is the possibility that the UV from HZ~Her are shielded
on their way to the X-ray emission region, but not on their way to our line
of sight. The 864~\AA \ photons which photoexcite \ion{Si}{13}
are particularly vulnerable since they are above the Lyman edge.
Such a shielding material could
be the accretion disk itself, in which case the origin of the recombination lines
in the disk atmosphere would be proven as well.
Previously, no distinction could be made between
the case of disk coronal emission and emission from the illuminated companion 
\cite[]{herx1me}.

We base our arguments above in the fact that the primary source of UV emission in the 
binary system is HZ~Her and not the accretion disk.
This is known from the orbital phase variability.
The UV continuum increases gradually from
$\phi \sim 0.2$ until it peaks at orbital phase $\phi \sim 0.5$, and then decreases 
gradually until the 
illuminated face of HZ~Her comes out of view at $\phi \sim 0.9$ 
\cite[]{uv_disk_lines}.
The intermediate mass ($\sim 2.3$\Msun) and size of HZ~Her causes it
to produce copious UV emission by reprocessing of the neutron star X-rays. 
The accretion disk emission in the 1260--1630~\AA \ band,
which is identifiable from eclipse ingresses and egresses at $\phi = \pm 0.1$,
represents a small fraction ($\sim 1/10$ to $1/20$) of the peak
HZ~Her emission at that band \cite[]{uv_disk_lines}. 
However, the question remains of how much UV continuum from the disk
is visible to the disk atmosphere.  Although the observed UV from the
disk is fainter than that from the companion, that is partly because the
disk is seen edge-on. 

In sum, the weakening of any photoexcitation effects in the $R$ ratios in 
\ion{Mg}{11} and \ion{Si}{13} places a lower limit on the distance from HZ~Her to
the X-ray emitting region, or a requirement on shielding from the UV.  
These limits show that the X-ray lines are not produced on the illuminated 
surface of the companion, but instead arise in a high density region
that is likely associated with the accretion disk.

\subsection{Search for orbital variability}
\label{sec:var}

The X-ray light curve is flat at mid-orbit. The light curves for the total and Fe~K$\alpha$ 
line fluxes
are shown in Figure \ref{fig:lc}.  A Bayesian block analysis \cite[]{scargle}
on these light curves showed no deviations from constant flux with 99.9\% 
confidence.  The coverage is limited to $1/3$ of the orbit, so 
we cannot entirely rule out orbital variability in the spectrum. After all,
Her X-1 exhibits eclipses in the low state \cite[]{xteobs}.
Clearly, the variability of the Her X-1 emission is distinct from
that observed in the ADC source 4U~1822-37,
the latter likely arising on an illuminated disk bulge \cite[]{4u1822}.
The UV continuum emission from HZ~Her shows a $\sim 25$\% 
flux variation from $\phi = 0.5$ to $\phi = 0.65$,
peaking at mid-orbit, and the lines
of \ion{O}{5} and \ion{N}{6} decrease in strength $\Delta \phi = 0.2$ 
away from eclipses \cite[]{uv_disk_lines}. 
With better phase coverage than ours, \xmm UV flux and Fe K$\alpha$ flux measurements
show a clear orbital modulation \citep[see their Fig. 9]{zane}, which suggests that at least
some of the Fe  K$\alpha$ flux originates on the face of the companion.
Our data show that the Fe  K$\alpha$ flux was unusually flat
during our observation.

\section{XSTAR Photoionized Plasma Models and Emission Measure}
\label{sec:em}

The differential emission measure (DEM) maps the ionization distribution of the plasma.
It serves as a basis for testing astrophysical emission models and measuring elemental
abundance ratios. A DEM analysis is valid insofar as the spectrum is dominated by 
recombination emission, which occurs when the 
optical depth is such that the plasma does not destroy or enhance
the lines above their recombination values. The luminosity of a line
from level $u$ to level $l$ is given by
\begin{equation}
dL_{ul} = n_e n_{z,i+1} E_{ul} \eta_{ul} \alpha_{RR} dV
\end{equation}
where $n_{z,i+1}$ is the number density of an ion with atomic
number $Z$ and charge $+(i+1)$, $E_{ul}$ is the transition energy, 
$dV$ is the differential volume, $\alpha_{RR} = \alpha_{RR}(T)$ is the total 
radiative recombination rate in units of ${\rm cm}^3~s^{-1}$,
and $\eta_{ul}$ is the fraction of all the recombination photons that produce
line emission through the transition $u \to l$. 
This can be related to the $DEM = \int d(EM)/d(\log \xi)$
through the $d(EM) = n_e^2 dV$ available at each $\log \xi$, since:
\begin{equation}
dL_{ul} = \frac{ n_{z,i+1} }{n_e} E_{ul} \eta_{ul} \alpha_{RR} d(EM) =  \frac{ n_{z,i+1} }{n_e} E_{ul} S_{ul} d(EM)
\end{equation}
where the ionic density $n_{z,i+1} = A_Z n_H f_{i+1}$ is often expressed in terms of the 
elemental abundance $A_Z$, the proton density $n_H$, and the charge state 
fraction of the recombining ion $f_{i+1}$. 
The specific line power is defined as $S_{ul} = \eta_{ul} \alpha_{RR}$.
Thus, the total line luminosity is
\begin{equation}
\label{eq:dem}
L_{ul} = \frac{ A_Z n_H }{n_e} E_{ul} \int f_{i+1} S_{ul} \frac{d(EM)}{d(\log \xi)} d(\log \xi)
\end{equation}
This sort of DEM procedure was 
illustrated by \cite{liedahl99} and applied by \cite{sako}, and it
is analogous to what has been used in collisionally ionized plasmas as $d(EM)/dT$
\cite[]{kaastra}.

In the simplest analysis, we estimate 
the total emission measure $EM = \int n_e^2 dV$ for each ion, 
assuming each is formed at a single $\xi_o$ and $T_o$. 
The EM for every prominent line in the spectrum is shown
in Figure \ref{fig:em}, calculated with the Solar abundances from \cite{grev},
with the updated C and O values from \cite{coabun}.
We estimate the elemental abundance ratios by enforcing 
the continuity of the EM as a function of $\xi_o$.
The \ion{N}{7} and \ion{N}{6} line fluxes 
show a N over-abundance with respect to C, O, Ne, Mg, Si, S, and Fe.
The \ion{C}{6} line flux is obtained from the \xmm RGS spectra
obtained by \cite{herx1me}.
From Equation \ref{eq:dem}, it is useful to adopt 
$\epsilon \equiv f_{i+1} S_{ul}$ (in units of photons cm$^3$~s$^{-1}$) 
as a measure of the line emissivity from now on. This $\epsilon$ is
a function of $\xi$.
We calculate the $f_{i+1}$ in a grid of models
with $0.0 < \log \xi < 5.0$ and $\Delta \xi = 0.1$,
using XSTAR \cite[]{xstar}. We use the main-on spectrum 
as the ionizing continuum.
The $S_{ul}$ are calculated with HULLAC \cite[]{hullac}.
The $\log~T_o$ and $\log~\xi_o$ of formation
for the Ly$\alpha$ or He$\alpha$ lines
are defined as the average values of $\log~T$ and $\log~\xi$
weighted by $\epsilon$.
The XSTAR/HULLAC model temperatures of formation
span $32,500~{\rm K} < T_o < 6.7 \times 10^6$~K, for 
\ion{N}{6} through \ion{Fe}{26}, respectively. 
We calculate the recombination rate at a single $\xi_o$ and 
assume $f_{i+1}=0.25$. We use $f_{i+1}=0.25$ because the $\log \xi$-averaged, emissivity-weighted
value of $f_{i+1}$ is approximately equal to that value.
While this is just an approximation, it is more realistic than simply setting $f_{i+1}=1$.
The EM analysis in Figure \ref{fig:em} has the advantages of being simple
and of making the result independent of the plasma models; however, 
one sacrifices the accuracy of the EM distribution with $\xi$.

In principle, a more accurate analysis than the above
is obtained with ${\rm DEM} = d({\rm EM})/d(\log{\xi})$,  
by assuming that the DEM is constant in the $\xi$ range at which each ion
is formed. Again, we enforce continuity of the DEM as a function of
$\xi$. We use the aforementioned Solar abundances.
Figure \ref{fig:emcorr} shows that
the N abundance has a large excess, C is moderately depleted, and O is near 
the Solar value. A recalculation of the DEM with the \cite{wilms} Solar abundances establishes that
the N excess is just as large, while both the C and O depletion appear significant.
This stems from the fact that the Wilms et al.~abundances for both C and O 
are larger than Allende Prieto et al.'s by $\sim 62$\% and $\sim 74$\%,
respectively. 
The abundance pattern, and especially the N abundance, is clearly a signature of CNO cycle
products, as shown previously by
the \xmm RGS spectral analysis, where the DEM was parameterized as 
a power-law and the Wilms et al. abundances were used \cite[]{herx1me}. 
The uncertainties in the Solar abundances do not affect this conclusion,
since no significant O depletion is needed.
The CNO process consists of the rapid conversion of C to N and
the $10^3$ times slower conversion of O to N \cite[and references therein]{clayton}.
Evolutionary models of 2--12\Msun \ stars produce an increase in the
N abundance by factors of a few, with C depletion in the tens of percent and
O depletion of a few percent or less, after the H-burning phase is over \cite[]{stellar}.
Clearly the most noticeable signature will be the N abundance.
It can be seen in the DEM that the abundance ratios among Ne, Mg, Si, S, and Fe 
are close to the Solar values.
Figure \ref{fig:emcorr} shows that successive pairs of the H-like and
He-like ions of a given element indicate that the DEM decreases with $\xi$,
from $\log \xi \sim 1.5$ to $\log \xi \sim 2.4$.
There, the DEM turns over and flattens out.
The consequences of the DEM shape being rather flat are explored in
the next section.
This DEM is sensitive to the modeled thermal stability of the plasma, as will
be discussed in \S \ref{sec:stab}.

\subsection{Volume, Scale Height, and Density Diagnostics}
\label{sec:volden}

Consider the filling volume $V$ of each line emission region in the optically thin limit.
Note that the $DEM$ is rather flat, i.e. 
it varies by less than an order of magnitude (once abundances are
accounted for). 
Since $\xi \propto n_e^{-1}$, a flat DEM requires that $V \propto \xi^2$.
Therefore, the region emitting \ion{Fe}{26} fills $\sim 10^2$ times more volume
than the \ion{Si}{13}-emitting region, and $\sim 10^4$ 
more volume than the \ion{N}{6}-emitting region.
The large disparity in the volume filled by different ions can be interpreted in two ways.
The first way is consistent with a model in which 
\ion{Fe}{26} is associated with hot corona with
$\sim 10^4$ times the scale height of the
compact, flattened disk atmosphere which produces
\ion{N}{6}. A scale height ratio of the same order was calculated from the hydrostatic 
model atmosphere \cite[]{diskmodel}.
A second way is a scenario in which the coldest, densest part of the
plasma is concentrated in small clumps,
while the hottest, more diffuse part of the plasma is distributed in a large volume.

Combining plasma diagnostics, photoionization balance, and emission measure,
we constrain the size and scale height of the line-emitting corona, 
set robust density constraints, and we estimate optical depths.
For this purpose, we select \ion{Fe}{26}, \ion{Mg}{11}, and \ion{O}{8}.
For each of these ions, we plot the density $n_e$ as a function
of distance $r$ from the X-ray source. The emitting plasma is most likely to reside inside the 
Roche lobe of the neutron star.
The $n_e(r)$ limits drawn for \ion{Fe}{26} in Figure \ref{fig:fe26} are:
\begin{enumerate}
\item An upper limit on the density set by thermal and ionization balance.
If the density is too high, the plasma would be too cold to emit the line,
because a given line can only be emitted at a fixed range of $\xi = L / ( n_e r^2 )$.
Thus, $n_e < L / (\xi_{\rm min} r^2)$,
where $\xi_{\rm min}$ is defined such that 95\% of the
line emissivity is at $\xi > \xi_{\rm min}$.
We take $L$ to be the main-on luminosity, which
is reprocessed into line emission during the low state.
\item A lower limit on the density given by the line flux through the emission measure EM
and the maximum volume $V_{\rm max}$,
such that $n_e > \sqrt { EM / V_{\rm max} } = \sqrt { 3 ~EM / (4 ~\pi r^3) }$.
\end{enumerate}
Figure \ref{fig:fe26} includes the density locus for a spherical shell of Thomson depth 
$\tau \sim 0.05$, which is consistent with the above limits.
Such is the depth required to Compton scatter the main-on continuum 
and produce the continuum observed in the low state. 
This region of the corona extends to $\sim 10^{11}$~cm. 
The coronal size derived from a shallow component of an eclipse ingress 
is $r \gtrsim 6 \times 10^{10}$~cm: the ingress X-ray light curves exhibit a steep and fast
component and a slow and shallow component, also implying a core of X-ray emission
concentrated at $r \lesssim 10^{9}$~cm \cite[]{eclipses_choi}.
High resolution spectroscopy is allowing us to probe the extended region
the corona. 
For the \ion{Fe}{26} corona with $\tau \sim 0.05$ to be enclosed by the Roche Lobe, 
the scale height must be $h/r \sim 0.1$, or more conservatively, 
$0.01 \lesssim h/r \lesssim 0.3$. 

We set stringent density constraints for the \ion{Mg}{11} 
emission region, taking advantage of the \ion{Mg}{11} $R$ ratio being at its
critical value (see \S \ref{sec:helike} and Fig. \ref{fig:mgxi}). 
We calculate the same $n_e$ limits as for \ion{Fe}{26}.
In addition, we calculate the critical $n_e$ (in the low UV flux limit), and
the critical $r$ found from UV photoexcitation (in the low density limit).
These are based on the $R$ ratio, as discussed in \S \ref{sec:helike}.
This allows us to compare the derived density with the modeled density,
for the case where the plasma is inside the Roche lobe.
To obtain the model density at the Her X-1 main-on luminosity,
we interpolate between two disk atmosphere models
at $L_{\rm edd}$ and $0.1~L_{\rm edd}$, 
and we find that the density predicted by the atmosphere model 
($n_e = 3 \times 10^{13}$\pcmcu) is
consistent with that derived directly from the $R$ ratio. 
This provides support to the disk atmosphere model, 
because photoionization balance alone would allow one or two
order of magnitude variations on $n_e$, depending on
the optical depth of the plasma.
The agreement in $n_e$ also implies
that density rather than photoexcitation effects are driving the
$R$ ratio in \ion{Mg}{11}. 

The \ion{O}{8} diagnostics in Figure \ref{fig:oviii}
allow us to assign an optical depth to a given geometry.
This figure shows the two $n_e$ limits discussed previously.
In addition, we plot the density of the plasma for two simple geometries: 
a spherical shell and a flat pill-box or atmosphere.
We show two solutions with optical depths that are consistent with the density limits.
The case of the spherical shell requires larger 
edge depths (of $\tau_{\rm O VIII} \sim 3$) than the case
corresponding to a flat atmosphere, for which $\tau_{\rm O VIII} \sim 0.01$
in the face-on direction, as would be the case for a warped disk.
The values that we will obtain from the optical depth diagnostics in \S \ref{sec:odepth}
will favor the scenario of a flat atmosphere.

\subsection{Abundances from the (\ion{N}{6} + \ion{N}{7})/(\ion{O}{7} + \ion{O}{8} ) line ratio}
\label{sec:noratio}

Using the photoionized XSTAR and HULLAC models discussed in the beginning of \S \ref{sec:em}, 
we quantify the relationship between the $P\equiv$~(\ion{N}{6}+\ion{N}{7})/(\ion{O}{7}+\ion{O}{8}) 
line {\it photon flux} ratio and the N/O elemental abundance ratio. 
Using the $P$ ratio as an abundance diagnostic has the virtue 
of being independent of the EM and DEM analyses, relying instead on the XSTAR plasma 
models and the recombination rates obtained from the HULLAC atomic code. 
The XSTAR models are calculated over a grid of $\log_{10} \xi$ values that cover the
range where the lines of interest are produced. We assume that the line emission is dominated by the
radiative recombination process, as is shown to be the case for 
Her X-1 in \S\ref{sec:odepth}. 
Figure \ref{fig:emis} shows the calculated 
$E \equiv$~[$\epsilon($\ion{N}{6})+$\epsilon($\ion{N}{7})] /
[$\epsilon($\ion{O}{7})+$\epsilon($\ion{O}{8})]
line {\it emissivity} ratio as a function of $\xi$, not accounting for elemental
abundances ($\epsilon$ was defined earlier in \S \ref{sec:em}). 
Therefore, we have $ P = E ~(N/O)$, where $(N/O)$ is the fractional
abundance ratio between nitrogen and oxygen.
The $E$ and $P$ ratios do not depend on density.
At practically any $\xi$ at which the lines of interest are produced, $0.7 < E < 2.4$.
The latter values represent the worst case error for one assuming $P \sim (N/O)$.
The measured line ratio ranges from 
$P = 0.87 \pm 0.11$ \cite[]{herx1me}
to $P = 1.54 \pm 0.42$ (this work). Combining these numbers, we
obtain $({\rm N/O})/({\rm N/O})_\sun = 5.9 \pm 0.6$, using the 
Allende~Prieto et al.~(2002) and \cite{grev} Solar abundances as reference,
and $({\rm N/O})/({\rm N/O})_\sun = 9.2 \pm 1.0$ 
using the Wilms et al.~(2000) Solar abundances. The error bars on these
abundance ratios only include statistical errors.
Compare this to the $4 \lesssim ({\rm N/O})/({\rm N/O})_\sun \lesssim 8$ 
that can be estimated by enforcing continuity on the DEM in Figure \ref{fig:emcorr}.

\section{Quantifying optical depth}
\label{sec:odepth}

X-ray spectroscopic analysis allows us to measure or set limits
on the mean number of scatters of line photons, or to the line optical depth.
The photoelectric optical depth can also be constrained by the spectrum,
despite the absence of absorption edges.
These measurements can have important implications for the geometry
of the plasma. They also allow us to validate the previously
determined elemental abundance ratios.

\subsection{The mean number of scatterings for Ly$\alpha$ and Ly$\beta$ photons}
\label{sec:lyb}

The mean number of scatterings is constrained by the observed 
Ly$\alpha$/Ly$\beta$ ratios. The Ly$\alpha$/Ly$\beta$ ratio is a temperature
diagnostic in collisional plasmas. In contrast, this ratio depends very weakly 
on temperature for a recombining plasma, as we have verified
from the HULLAC recombination rates.
A Ly$\beta$ photon from a H-like ion can 
get resonantly absorbed by another ion of the same species (see Fig. \ref{fig:hlike}). There is 
a probability $\gamma = 0.12$ that the
absorbed Ly$\beta$ photon is converted to photons with lower energy 
(including an H$\alpha$). Otherwise, the absorbed Ly$\beta$ is re-emitted.
The probability of a Ly$\beta$ photon to survive $N$ line scatterings is:
\begin{equation}
\label{eq:simple}
P = (1-\gamma)^N =  0.88^N,
\end{equation}
and therefore the Ly$\beta$ photons 
cannot scatter more than a few times before getting destroyed.
As the number of scatterings increases, the
Ly$\alpha$/Ly$\beta$ intensity ratio increases with it,
because Ly$\alpha$ photons do not get destroyed in line scatterings ($\gamma = 0$).
None of the observed Ly$\alpha$/Ly$\beta$ line ratios 
shown in Table \ref{tab:lyb} show evidence for multiple line scatterings,
except for \ion{O}{8}. Since we deal with a large ensemble of photons, 
the number related to the line ratio is the mean number of scatterings $\langle N \rangle$,
instead of $N$ above.

The observed Ly$\alpha$/Ly$\beta$ ratio 
allows us to constrain the mean number of scatterings 
of Ly$\beta$ photons directly, and of Ly$\alpha$ photons by inference.
If $\gamma > 0$, the mean number of scatterings $\langle N \rangle$ for a slab geometry 
satisfies \cite[]{hummer} :
\begin{equation}
\label{means}
\langle N \rangle = \frac{1 - \gamma}{\gamma} \ \frac{E_L}{E_G}
\end{equation} 
where $E_L$ is the line energy lost to resonant absorption and $E_G$ is the total
line energy generated in the gas column.
There are two relevant cases for the upper limit of $\langle N \rangle$.
In the case $\gamma = 0$, photons do not get destroyed (as for Ly$\alpha$), then
$\langle N \rangle$ does not have an upper bound, and Equation \ref{means} does not apply.  
In the case $\gamma > 0$ (as for Ly$\beta$),
$\langle N \rangle$ is bounded above (by $\langle N \rangle < 7.33$).
as $ E_L/E_G \to 1$ in Equation \ref{means}.

We first use Equation \ref{means} for \ion{O}{8}.
We detected 7 counts of the \ion{O}{8} Ly$\beta$ line
in the 15.995~\AA \ to 16.015~\AA \ range. Given the
continuum and background level of 3.0 counts, this represents a detection
significance of 95\%, as derived from a Poisson distribution. 
The caveat is that this is only valid if the background is uniform in $\lambda$,
and it may be subject to systematics.
Since this is not
a line search (i.e. the line energy is fixed), this may be a positive detection, but it
needs confirmation.
Taking Ly$\beta$ at face value, we derive
$E_L/E_G < 0.9$ for \ion{O}{8} Ly$\beta$ with 90\% confidence, and obtain
$\langle N \rangle < 6.6$.
The detected line is centroided at 16.005~\AA \ (\ion{O}{8} Ly$\beta$), and it is 
one MEG-resolution FWHM away from \ion{Fe}{18} at 16.023~\AA.

To infer the mean number of scatterings of Ly$\alpha_1$ from 
those of Ly$\beta$, we need to relate
$\langle N \rangle$ to $\tau$, the mean line optical depth. \cite{hummer} 
performed such a calculation for a slab geometry with a uniform source of line photons, $\gamma =0$, 
and Voigt parameters of $4.7 \times 10^{-2}$ to 
$4.7 \times 10^{-4}$. They obtained $\langle N \rangle / \tau \sim 0.55$--0.6
for $\tau = 1$, of $\langle N \rangle / \tau \sim 0.60$--0.65 for $\tau = 10$, 
and $\langle N \rangle / \tau \sim 0.73$--0.85 for $\tau = 100$. 
The Voigt parameter is the ratio between the natural broadening and the velocity 
broadening of the line.  Define $c_\tau = \langle N \rangle / \tau$. 
To generalize this result for $\gamma > 0$, note from Equation
2.15 in \cite{hummer} that $( 1-\gamma )$ can be factored out of the expression 
for $\langle N \rangle$, such that $\langle N \rangle = c_\tau \tau (1-\gamma )$
is applicable in that case, with $c_\tau$ and $\tau$ as before. Note $c_\tau = c_\tau(\tau)$ 
is a function of $\tau$.

The mean line optical depth $\tau$ is proportional to the oscillator
strength $\tau \propto f_{\rm osc}$.
Therefore, we obtain $\langle N \rangle \propto c_\tau f_{\rm osc} (1-\gamma)$,
from which we estimate $\langle N \rangle \sim 30$ for \ion{O}{8} Ly$\alpha_1$ photons. 
A similar number can be obtained through the more simplistic Equation \ref{eq:simple}.
We ignored the continuum optical depth. The effect of a nonzero continuum depth 
is to decrease $\langle N \rangle$ below the derived values.
Up to this point, this result depends on the detection of the Ly$\beta$ line, but there
is another way to deduce $\langle N \rangle$ for \ion{O}{8} Ly$\alpha_1$ photons.

We set an upper limit on $\langle N \rangle$ for \ion{O}{8} Ly$\alpha_1$ without 
relying on \ion{O}{8} Ly$\beta$, by using the \ion{Ne}{10} lines. 
This is valid since both ions are produced at a 
largely overlapping ionization parameter range. 
We find that $\langle N \rangle \lesssim 2$ for
\ion{Ne}{10} Ly$\beta$ photons (at 90\% confidence).
Using the $c_\tau$ once again, this implies
$\langle N \rangle \lesssim 9$ for \ion{Ne}{10} Ly$\alpha_1$.
The EM derived from \ion{Ne}{10} and \ion{O}{8} are the within 12\% of each other,
consistent with being equal given the statistical uncertainties.
Scaling with the \cite{wilms} or \cite{coabun} Solar abundances, we deduce
$\langle N \rangle \lesssim 69$ for \ion{O}{8} Ly$\alpha_1$.
Scaling with CNO abundances, $\langle N \rangle$ is smaller than or equal
to the latter limit.
Both are consistent with the $\langle N \rangle \sim 30$ value for \ion{O}{8} Ly$\alpha_1$ 
which was derived above.

In contrast, the Ly$\alpha$/Ly$\beta$ ratios of \ion{Mg}{12} and \ion{Si}{14} on
Table \ref{tab:lyb} may show evidence for enhancement of Ly$\beta$ due to
photoexcitation effects at moderate column densities. Photoexcitation
of ions by X-ray continuum photons, followed by radiative
decay, can dominate the emission mechanism. This process, also referred to as
resonant scattering, can produce a characteristic Ly$\alpha$/Ly$\beta$ ratio \cite[]{ali}.
At low column densities, both lines are unsaturated, resonant scattering
dominates over recombination emission,
and the Ly$\alpha$/Ly$\beta$ ratio simply depends on the oscillator strengths.
At moderate column density,
the Ly$\alpha$/Ly$\beta$ ratio is reduced below the
low column value, when the Ly$\alpha$ line reaches saturation but the Ly$\beta$ line
does not.
The evidence for photoexcitation of \ion{Mg}{12} and \ion{Si}{14} at the
moderate column density regime is a $\sim 2 \sigma$ result, so we do not discuss
this scenario further.
At high column densities, recombination emission dominates over resonant scattering,
because resonant scattering has completely saturated while the emission measure
keeps growing with the column,
and the Ly$\alpha$/Ly$\beta$ ratio is determined by the ratio of
the recombination efficiency of the lines,
which happens to be similar to the oscillator strength ratio.
In this section we have expanded upon the \cite{ali} picture, showing
that at high line optical depths, the Ly$\alpha$/Ly$\beta$ ratio
can increase above its pure recombination value due to the destruction
of the Ly$\beta$ line.

\subsection{The optical depth of the $r$ line in He-like ions}
\label{sec:hea}

The value of the $G = (i+f)/r$ ratio depends on whether the plasma
is collisionally ionized or photoionized, but it can also can be used to 
quantify the mean optical depth of the $r$ line.  The $G$ ratio depends 
on optical depth for three reasons: 1)
the $r$ line has much larger oscillator strength than either the $i$ or
$f$ lines; 2) the recombination rates of the $i,f$ lines are
larger than those of the $r$ line, because the $i,f$ lines
have larger statistical weight than the $r$ line; 
and 3) the $G$ ratio depends very weakly on temperature in a photoionized plasma.
This diagnostic has been quantified by
\cite{ali} and \cite{patrick}. 
In the limit of small line optical depth, we have
$G \sim 0.3$, because the $r$ line is enhanced relative
to the $f$ and $i$ lines by continuum X-rays that photo-excite the ground state 
\cite[]{patrick}. In the small line depth regime, the emission
is dominated by resonant scattering over recombination
emission.  In this regime, the line spectrum is dependent on the spectral 
energy distribution.
In the limit of large line optical depth, $G$ reaches an asymptotic value of 4.5
\cite[]{porquet}, which depends very weakly on the the $T$ values derived from 
photoionization equilibrium, and recombination emission dominates over 
resonant scattering.

The observed $G$ ratios shown in Table \ref{tab:helike}
are consistent with the limit of large line optical depth. By using the lower bounds on $G$,
we can set lower limits on the mean depth of the $r$ lines as defined by
\cite{patrick}, of $\tau_r \gtrsim 100$ for $G > 4$ (the maximum value applicable to
\ion{Ne}{9} and \ion{O}{7}), $\tau_r \gtrsim 20$ for $G > 3$ (applicable to \ion{Mg}{11}), 
and $\tau_r \gtrsim 10$ for
$G > 2$ (applicable to \ion{Si}{13}), assuming a continuum power-law index of $\alpha = 1.0$. 
The observed main-on index is $\alpha = 0.85$, but $G$ depends weakly on $\alpha$.

\subsection{An X-ray atmosphere illuminated at a small grazing angle
has anisotropic line depths} 

If the plasma geometry is anisotropic, the line optical depth will depend 
on direction. A first special direction vector
corresponds to the path of the X-ray continuum photons from the X-ray pulsar, since
this radiation is photoionizing the plasma. The second
special direction is that which minimizes the optical depth, since photons
will tend to escape from the plasma that way.
Based on this, we define two relevant depths: an ``illumination'' optical depth $\tau_i$, and
an ``escape'' optical depth $\tau_e$, which are well-defined and not equal if the 
plasma distribution is illuminated along a direction where
the optical depth is not minimized. 
The illumination depth $\tau_i$ determines how much of the 
continuum radiation is absorbed, and therefore how much is reprocessed into
resonant and recombination emission lines.
On the other hand, once a line photon is produced, 
it is more likely to escape in the direction where
the optical depth is minimized than through other directions, and this
corresponds to the escape depth $\tau_e$.

The He$\alpha$ $G$ ratio is sensitive to $\tau_i$,
because $G$ depends on the ratio between the resonant scattering and
recombination line fluxes, as mentioned in the previous section.  
Further scattering of the $r$ line after it is produced does not destroy it
unless the continuum depth is significant. 
On the other hand, the Ly$\alpha$/Ly$\beta$ ratio is sensitive to both $\tau_i$
and the escape depth $\tau_e$. 
Again, resonant line absorption of Ly$\alpha$ or Ly$\beta$ produces a dependency on 
$\tau_i$, because both photons resonant scatter strongly.
In addition, the Ly$\beta$ photon can be destroyed and converted to Ly$\alpha$
if it scatters too many times inside the plasma, and this is why the 
dependence on the escape depth $\tau_e$ arises.

The flat, geometrically thin shape of the model disk atmosphere 
is anisotropic, and the atmosphere is illuminated by the
neutron star at a very shallow grazing angle that
was determined from self-consistent calculations \cite[]{diskmodel}. 
This geometry is shown schematically in Figure \ref{fig:depth}.
In the atmosphere model, the calculated grazing angle is such that
illumination depth is 25 times larger than the escape depth at
$r = 10^{11}$~cm, and therefore the
observed He$\alpha$ $r$ line depth is expected to be 
much larger than the Ly$\beta$ depth. 
Sample vertical column densities in the disk atmosphere model are
$N_{\rm O VIII} = 5 \times 10^{17}$\pcmsq, $N_{\rm O VII} = 2 \times 10^{17}$\pcmsq, and
$N_{\rm Si XIII} = 1.6 \times 10^{16}$\pcmsq.
These columns imply escape line depths of $\tau_e \sim 46~(150~{\rm km~s^{-1}}/v)$ 
for \ion{O}{8} Ly$\alpha_1$, $\tau_e \sim 53~(150~{\rm km~s^{-1}}/v)$ 
for \ion{O}{7} $r$, and of $\tau_e \sim 1.4~(150~{\rm km~s^{-1}}/v)$ 
for \ion{Si}{13} $r$, where $v$ a fiducial velocity width that includes thermal and turbulent
components. Setting $v$ to the thermal value at $kT = 25$~eV for \ion{O}{8} yields
$\tau_e = 178$ for Ly$\alpha_1$, and setting 
$kT = 3.5$~eV for \ion{O}{7} yields $\tau_e = 544$ for He~$r$.
Note that $\tau_i \sim 25 \tau_e$ for the disk model geometry, i.e. for
\ion{Si}{13} we get $\tau_i = 36~(150~{\rm km~s^{-1}}/v)$.
These are the expected line depths from the disk atmosphere model, modulo the
assumptions made about the gas kinematics.

Compare the above values with the optical depth measurements from
Ly$\alpha$/Ly$\beta$ that yielded 
$\langle N \rangle \sim 30$ for \ion{O}{8} Ly$\alpha_1$, which is
equivalent to $\tau \sim 46$ given the calculated $\langle N \rangle /\tau$ ratios
and Voigt parameters from \S \ref{sec:lyb}. This optical depth matches the model at
the chosen 150\kmps velocity width. If the \ion{O}{8} Ly$\beta$ line detection is not real,
from \ion{Ne}{10} it follows that for \ion{O}{8} Ly$\alpha_1$ still $\tau \lesssim 100$. 
This would simply require a smaller velocity width than above, so there is parameter
space available to achieve consistency of the data with the disk atmosphere model in this case.
For the case of \ion{O}{7}, an upper limit $\tau_r > 100$ was obtained from the
$G$ ratio, which is easily accommodated by  $\tau_i \sim 1300~(150~{\rm km~s^{-1}}/v)$
for the illumination depth. This large $\tau_i$ leaves plenty of parameter space
to be consistent with the other lower limits on $\tau_r$ set for
\ion{Ne}{9}, \ion{Mg}{11}, and \ion{Si}{13}. For example, \ion{Si}{13}
was measured to have $\tau_r \gtrsim 10$, which is consistent
with the atmosphere model $\tau_i = 36$.
For reference, the model
escape depth for the continuum is $\tau_e{\rm (O VIII)} = 0.05$,
an illumination depth $\tau_i{\rm (O VIII)} = 1.2$.

Our results above depend on the estimated kinematics, and therefore are only
approximate consistency checks.
Since the velocity field in the disk is produced by both Keplerian and turbulent
motions, $\tau_e$ will in fact be highly anisotropic, and it will depend on the
photon momentum vector relative to the disk velocity field.  To self-consistently derive 
a turbulent velocity from the Ly$\alpha$/Ly$\beta$ ratio, one 
requires a calculation of the line transfer which accounts for this velocity field.

\subsection{RRC/RR ratios and electron temperature} 

The electron temperature measured from the \ion{Ne}{9} RRC
is $kT = 7 \pm 3$~eV, or $T = 81000 \pm 35000$~K. This was
measured by fitting the semi-Maxwellian profile of the
\ion{Ne}{9} RRC at $\lambda = 10.37$~\AA, the
brightest RRC in the spectrum, detected at the 5$\sigma$ level 
(see Fig. \ref{fig:megheg}). 
The XSTAR plasma code and the HULLAC recombination rates yield
emissivity-weighted average temperatures $kT_o$
which are shown on Tables \ref{tab:lyb} and \ref{tab:rrc}.
The $T_o$ for \ion{Ne}{9} is
in good agreement with the measured $T$. 
The significance of detection for other
RRCs obtained with the Cash statistic ranges from the 
2$\sigma$ to the 5$\sigma$ 
level (see Table \ref{tab:lines}). To fit the RRCs, we rebin
the spectra to bin sizes $\Delta \lambda$ of 0.05~\AA \ to 0.09~\AA \
(coarser than the bin size used for fitting lines).
While most RRCs do not have
sufficient statistics for an accurate $T$ measurement, $T$ is left
as a free parameter and the best-fit values are consistent with 
$T_o$ for most cases.

In Table \ref{tab:rrc}, we show the observed  
RRC to RR (radiative recombination lines) 
flux ratios, compared to the HULLAC calculations,
in the recombination-dominated case, for both a single temperature model
and for the disk atmosphere model.
The RRC/RR ratios for \ion{O}{7}, \ion{O}{8}, and \ion{Ne}{10}
are significantly weaker than expected for the single-$T$ model,
but agreement is generally better for the disk atmosphere model, which
accounts for a broad distribution of temperatures $2 < kT < 860$~eV.
It is therefore likely that
the RRCs are composed of multiple temperatures, with the
low-$T$ component being observed more readily than the
high-$T$ component. The high-$T$ RRC component, which usually 
dominates the recombination flux,
is too broad to be separable from the continuum. In this
favored scenario, the weak RRCs are simply an indication of the multi-$T$
nature of the plasma. 

\subsection{Verification of CNO abundances: the line optical depths are
not large enough to convert \ion{O}{8} Ly$\alpha$ to \ion{N}{7} Ly$\alpha$ }
\label{sec:cno}

We test for the presence of the X-ray Bowen fluorescence mechanism by which a
\ion{O}{8} Ly$\alpha$ line becomes a \ion{N}{7} Ly$\alpha$ line, as calculated
by \cite{sakorad}.  We determine whether this X-ray Bowen fluorescence effect
can account self-consistently for the apparent 
excess intensity observed in \ion{N}{7} Ly$\alpha$, \ion{N}{6} $i$ and $r$, and
the \ion{N}{7} RRC with respect to other lines, such as the \ion{O}{8} Ly$\alpha$
and \ion{O}{7} $i$ and $r$. We find that:
\begin{enumerate}

\item The X-ray Bowen fluorescence mechanism is unlikely
to enhance \ion{N}{6} He$\alpha$ emission in
the same proportion as \ion{N}{7} Ly$\alpha$.
This would require line opacities to overlap 
in energy space under nearly identical conditions as those found for 
\ion{N}{7} Ly$\zeta$ and \ion{O}{8} Ly$\alpha_2$, whose energies 
are 1 Doppler width away at $T=50~$eV.
We checked for coincidences between the energies of the brightest lines of 
abundant elements in the 22.46 to 26~\AA \ band and the resonance line energies 
of \ion{N}{6}. We did not find line overlaps closer than 46 Doppler widths at $T = 50$~eV,
and the number of Doppler widths is likely to be higher since models show 
a peak \ion{N}{6} He$\alpha$ flux at $T=2.8$~eV. 
As shown in the HETGS spectrum in Fig. \ref{fig:megon}, 
as well as from the \xmm RGS spectrum \cite[]{herx1me}, 
{\it both} \ion{N}{6} He$\alpha$ and \ion{N}{7} Ly$\alpha$ indicate
a N over-abundance. 

\item The measured mean number of scatterings of
\ion{O}{8} Ly$\alpha$ and Ly$\beta$ photons is two
orders of magnitude lower than required for the X-ray Bowen fluorescence effect to take place.
The measured line ratio between
\ion{O}{8} Ly$\alpha$ and Ly$\beta$ depends on  $\langle N \rangle$,
due to the destruction of Ly$\beta$ or its conversion to Ly$\alpha$. 
We derive $\langle N \rangle \sim 30$ for \ion{O}{8} Ly$\alpha_1$. This was
verified via the same ratio in \ion{Ne}{10}, which scales to
$\langle N \rangle \lesssim 69$ for \ion{O}{8} Ly$\alpha_1$ with Solar or CNO abundances
(see \S \ref{sec:lyb}).
Compare this to the $\langle N \rangle \sim 5 \times 10^3$ scatterings needed for
the \ion{N}{7} enhancement \cite[]{sakorad}. 

\item The detection of the \ion{N}{7} 
RRC with \xmm and \chandra provides
additional evidence for a N/O overabundance.
At the escape line depths implied by the measured Ly$\alpha$/Ly$\beta$ ratios,
the RRC fluxes are unaffected by optical depth effects. 
The \ion{N}{7} RRC detected with \xmm RGS \cite[]{herx1me}
at 10$\sigma$ has flux $(6.5 \pm 2.8) \times 10^{-5}$\phpcmsqps,
with 90\% confidence errors, and HETGS detected
it at 3.5$\sigma$ with flux  $(18 \pm 5) \times 10^{-5}$\phpcmsqps,
with 68\% confidence errors (Table \ref{tab:lines}).
These fluxes are larger than the $\lesssim 1 \times 10^{-5}$\phpcmsqps
expected from Solar abundances and the fluxes of \ion{O}{7} RRC and \ion{O}{8} RRC.  

\item An analysis of the UV line emission from Her X-1 using
the same sort of illuminated disk model concluded that
N was over-abundant in HZ~Her, with a N/C abundance ratio of $\sim 2$
\cite[]{raymond}, which implies a N overabundance relative to Solar of $\sim 6$.
This shows that both UV and X-ray disk emission
models yield consistent results.

\item The velocity shear in the disk atmosphere and
corona increases the line escape probability, which would
decrease $\langle N \rangle$ below the static values calculated by
\cite{sakorad} for a given column density. This suppresses X-ray Bowen Fluorescence.


\end{enumerate}

We conclude that the large nitrogen over-abundance interpretation of the Her X-1 spectrum 
by \cite{herx1me} is valid and cannot be self-consistently 
explained by the X-ray Bowen fluorescence effect.
The large nitrogen over-abundance implies that significant CNO 
processing occurred in a massive star in the system.
New transfer calculations are required
to improve the optical depth
and dynamical constraints on the accretion flow.
Such calculations need to include
thermal and ionization balance, stratification,
and gas orbital kinematics in order
to calculate line transfer more accurately than existing models.

\section{Comparison with an accretion disk atmosphere and corona model}
\label{sec:modcomp}

\begin{figure}
\epsscale{1.}
\plotone{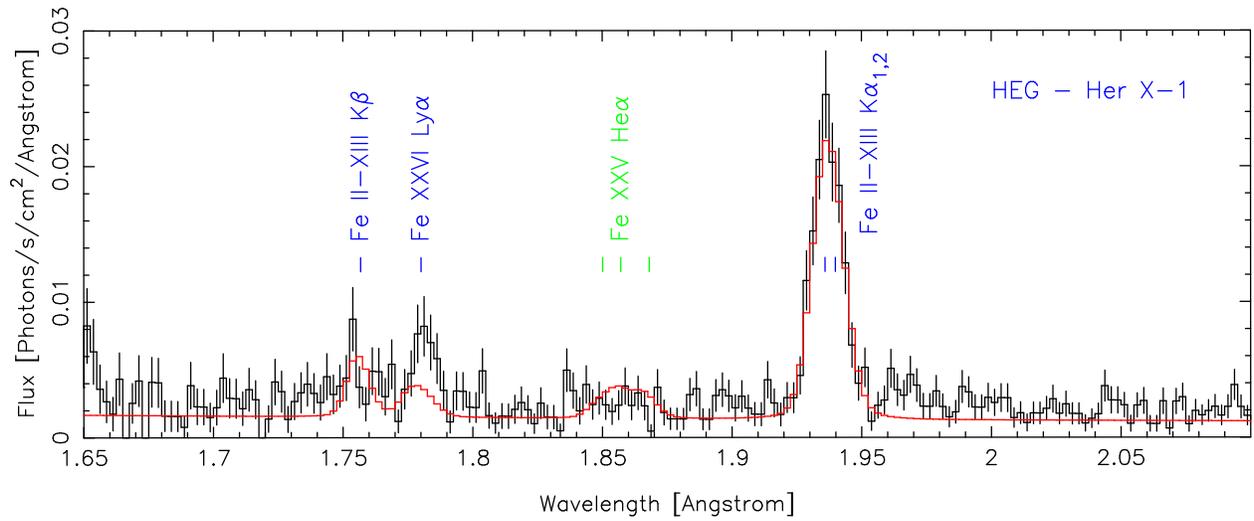}
\caption{HEG spectrum, showing both \ion{Fe}{1}-\ion{Fe}{13} K$\alpha$ and K$\beta$ fluorescence
lines, as well as the \ion{Fe}{26} Ly$\alpha$ radiative recombination line.
The disk corona and atmosphere model is over-plotted in red, except for 
the fluorescence lines, which were fit independently. 
The green label indicates that the line was produced in the model
but not detected.
\label{fig:fe}}
\end{figure}

\begin{figure}
\epsscale{1.}
\plotone{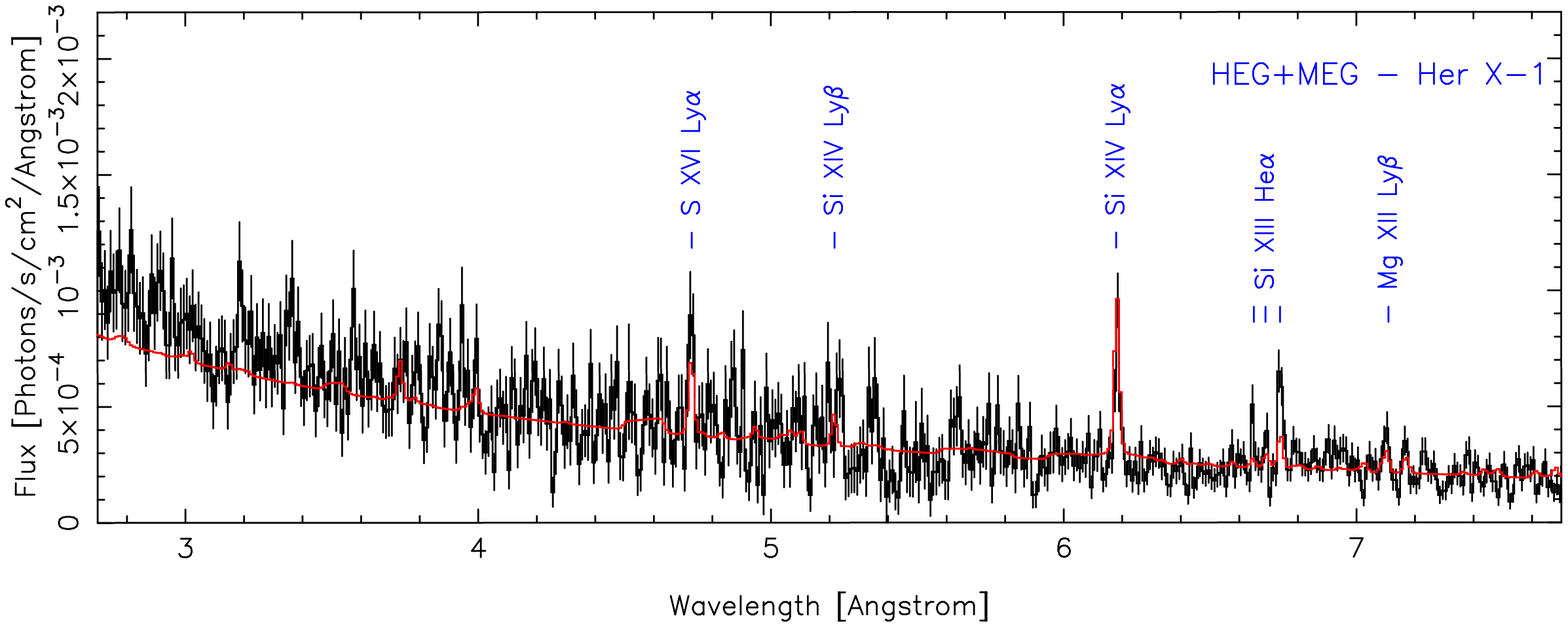}
\plotone{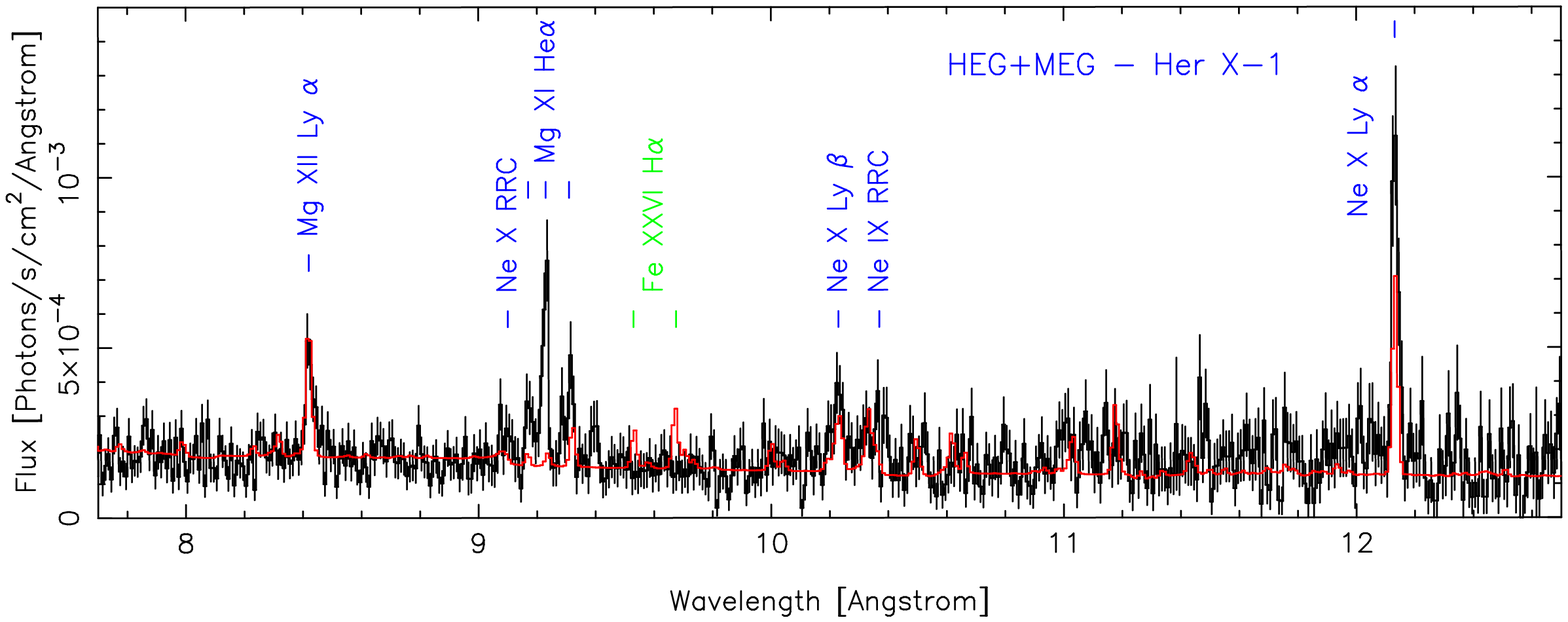}
\caption{Added HEG and MEG spectra.
The disk corona and atmosphere model is over-plotted in red.
The green label indicates that the line was produced in the model
but not detected.
\label{fig:megheg}}
\end{figure}

\begin{figure}
\epsscale{1.}
\plotone{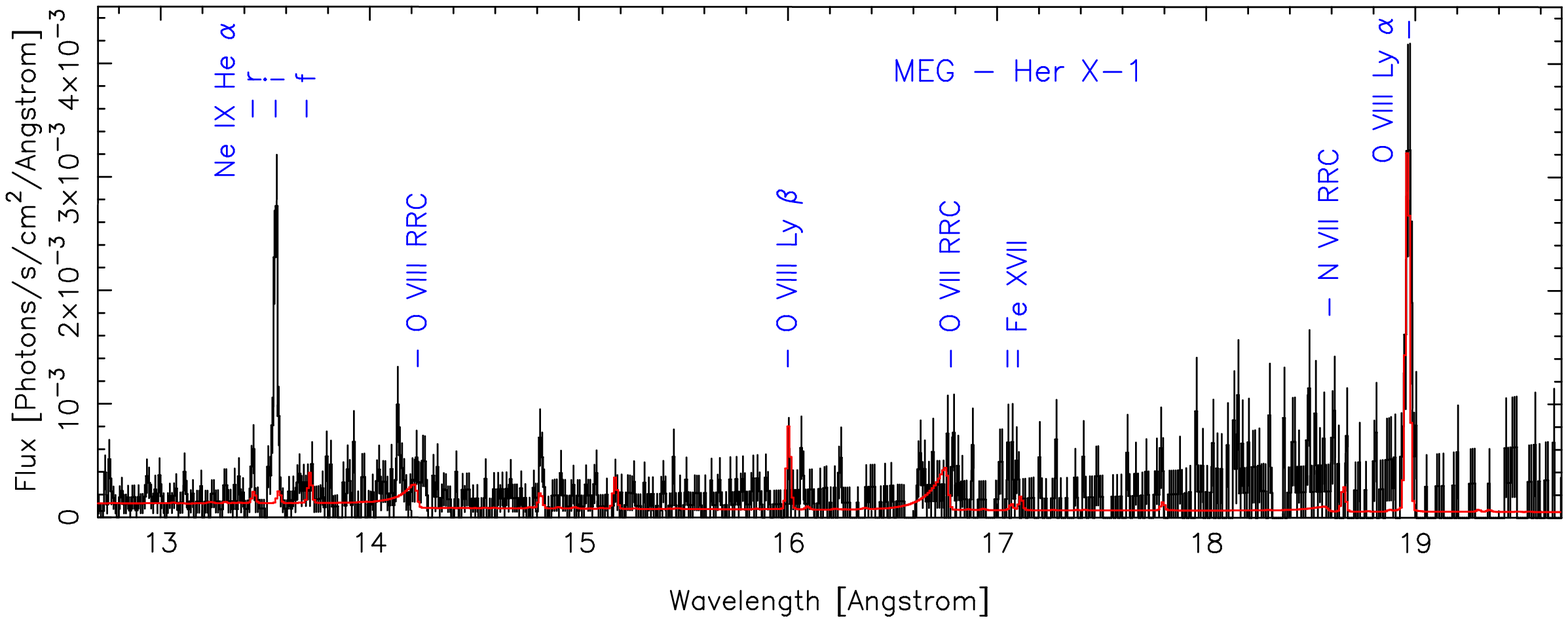}
\plotone{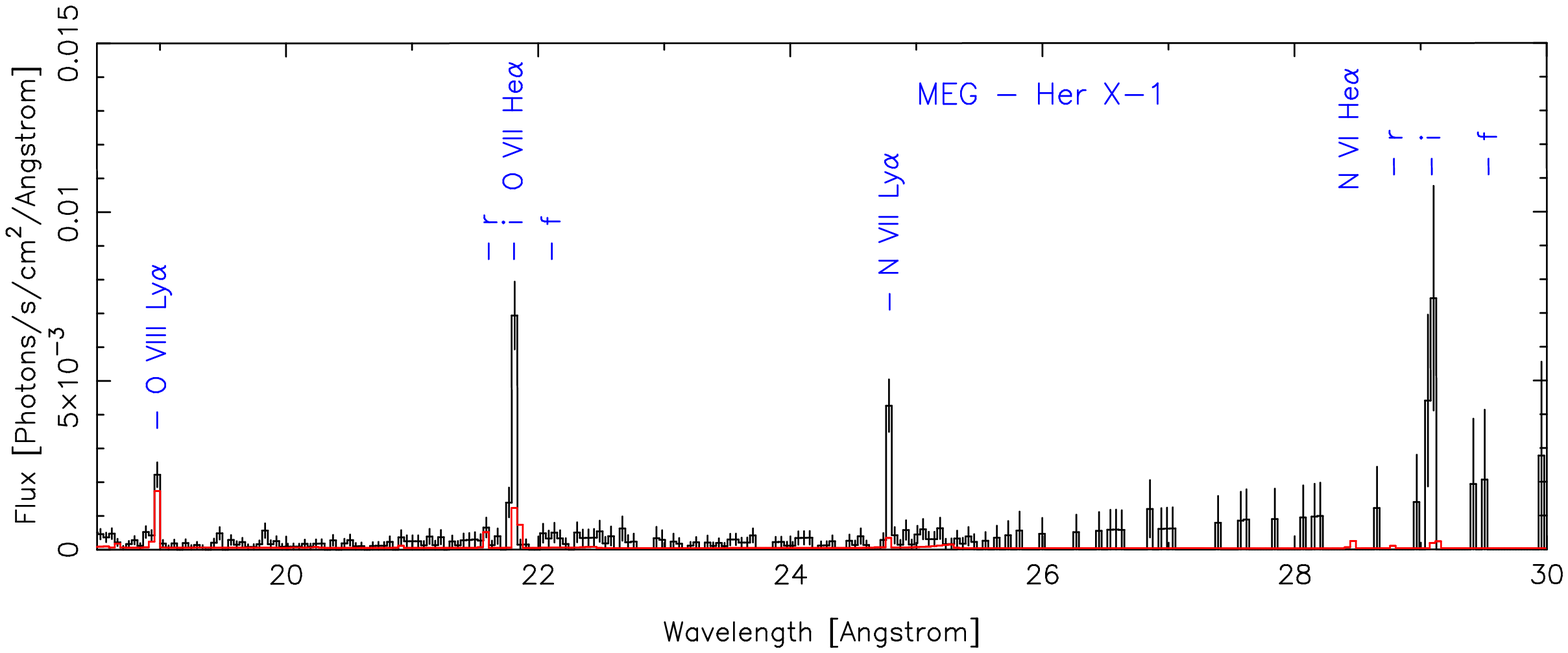}
\caption{MEG spectra which show that the nitrogen lines are quite
bright relative to the oxygen lines, as observed with \xmm \cite[]{herx1me}.
The disk corona and atmosphere model is shown in red.
\label{fig:megon}}
\end{figure}

\begin{figure}
\epsscale{0.4}
\plotone{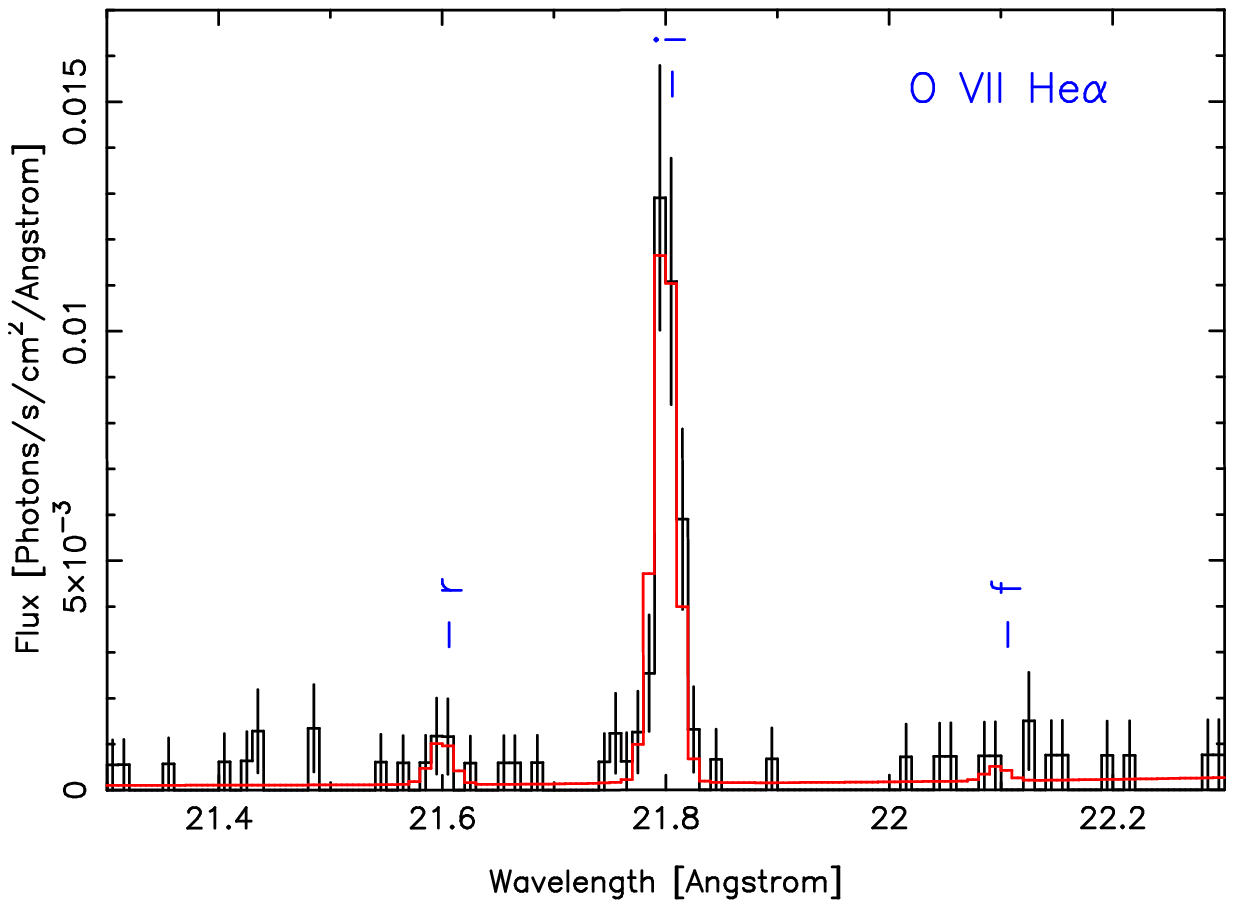}
\plotone{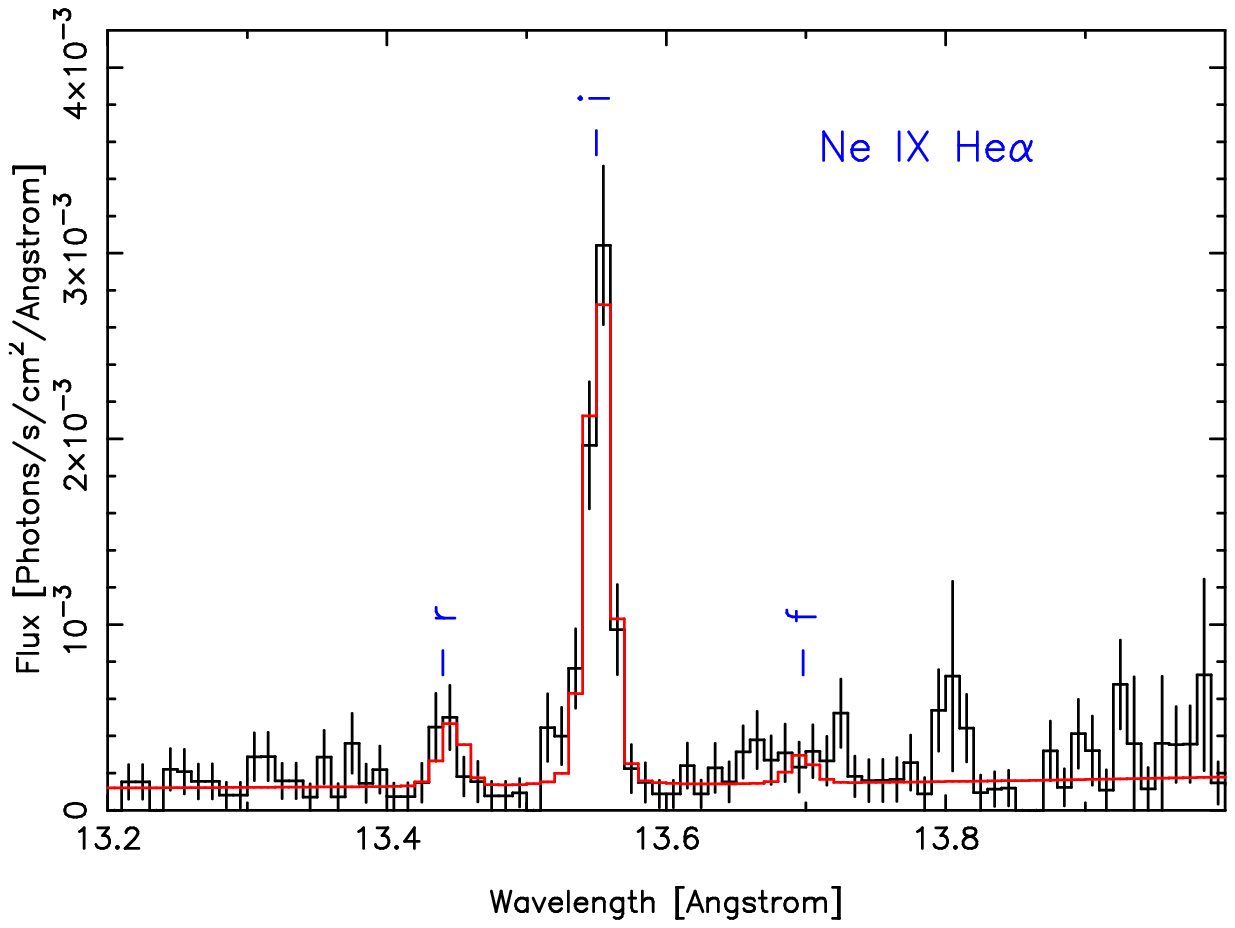}
\plotone{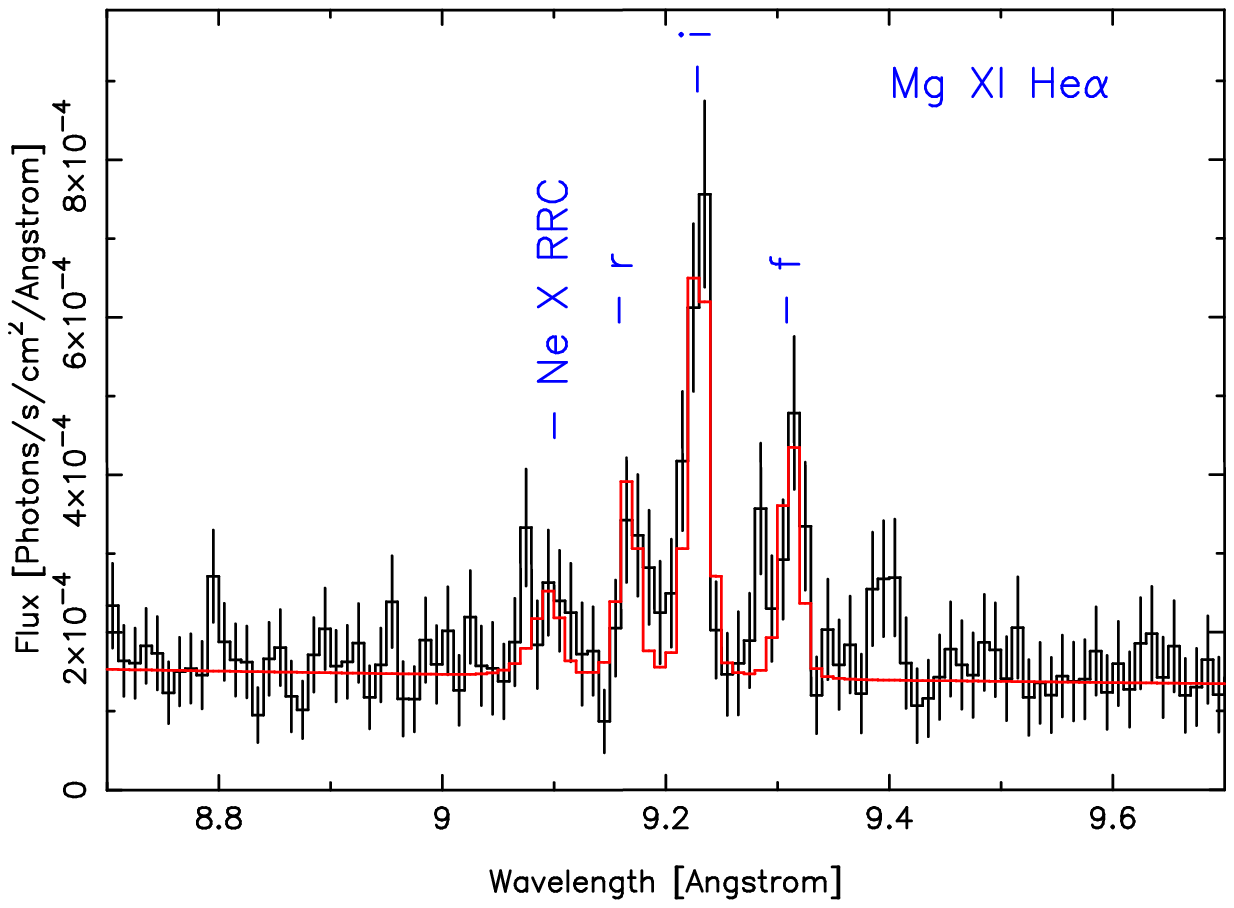}
\plotone{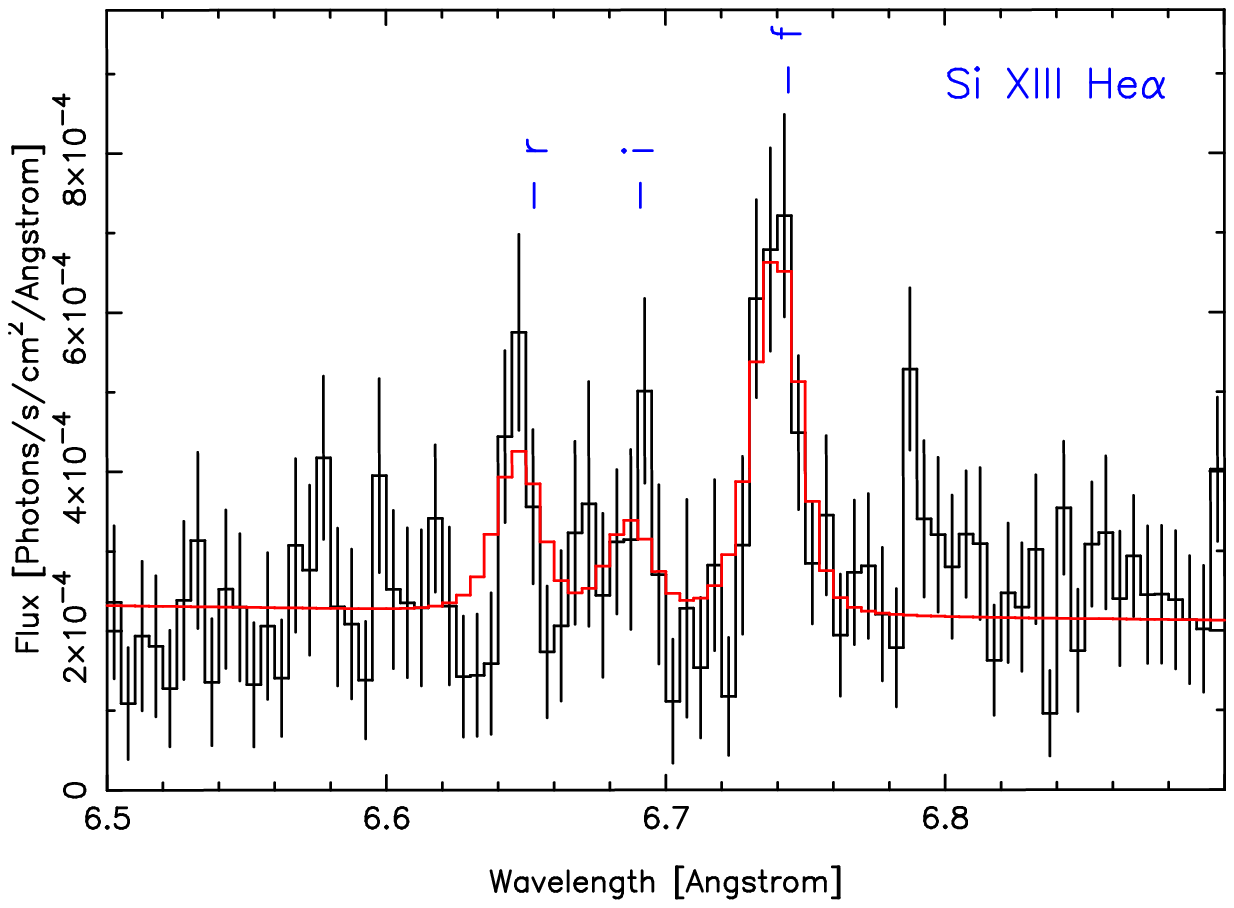}
\caption{He-like triplet lines for \ion{O}{7}, \ion{Ne}{9}, \ion{Mg}{11}, and
\ion{Si}{13}. The line ratios evolve smoothly from 
\ion{O}{7} and \ion{Ne}{9} with $R \sim 0$, to
\ion{Si}{13} with $R = 3 \pm 1$.
The $R$ ratio of \ion{Mg}{11} has a critical value $R = 0.51 \pm 0.12$,
which can provide a precise measure of the density.
The Gaussian line fits (not from the disk model) are over-plotted in red.
\label{fig:triplet}}
\end{figure}

\begin{figure}
\epsscale{1.}
\plottwo{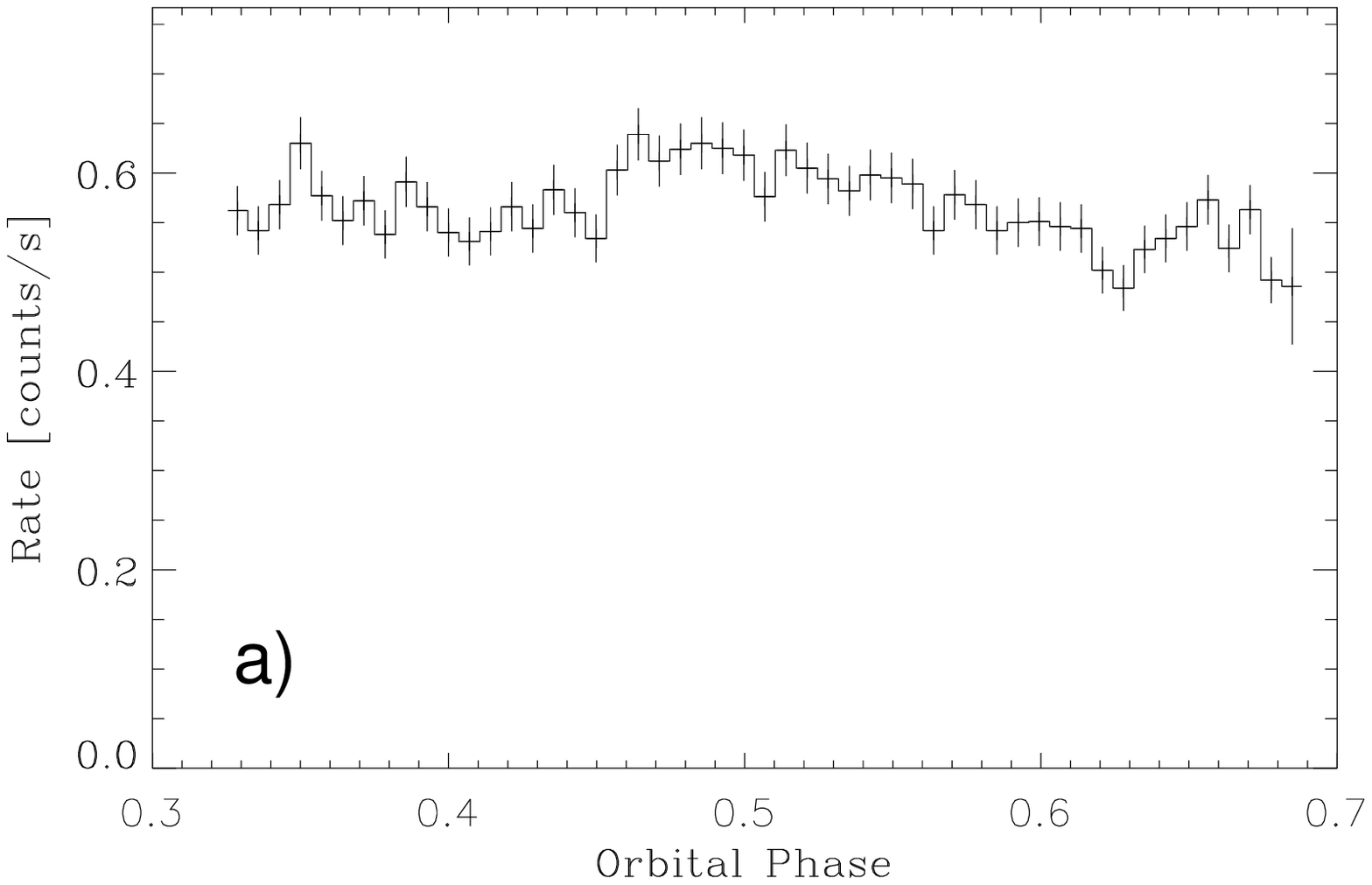}{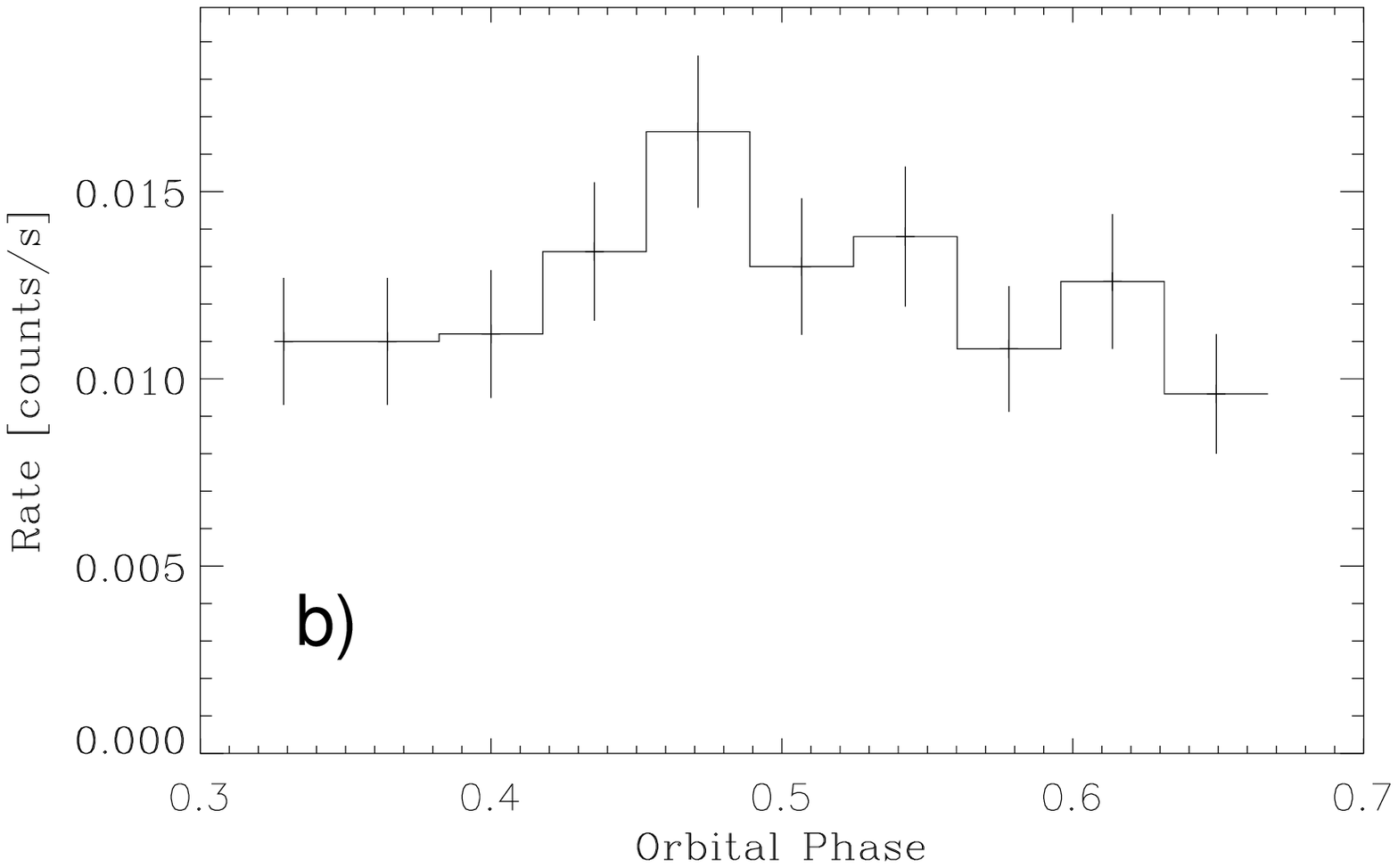}
\caption{Light curves of: {\it a)} the total HETGS dispersed count rate with $\lambda > 1.5$~\AA,
and {\it b)} the \ion{Fe}{1}-\ion{Fe}{13} K$\alpha$ line HETGS count rate.
The orbital period is 1.7~days.
\label{fig:lc}}
\end{figure}

\begin{figure}
\epsscale{1.}
\plotone{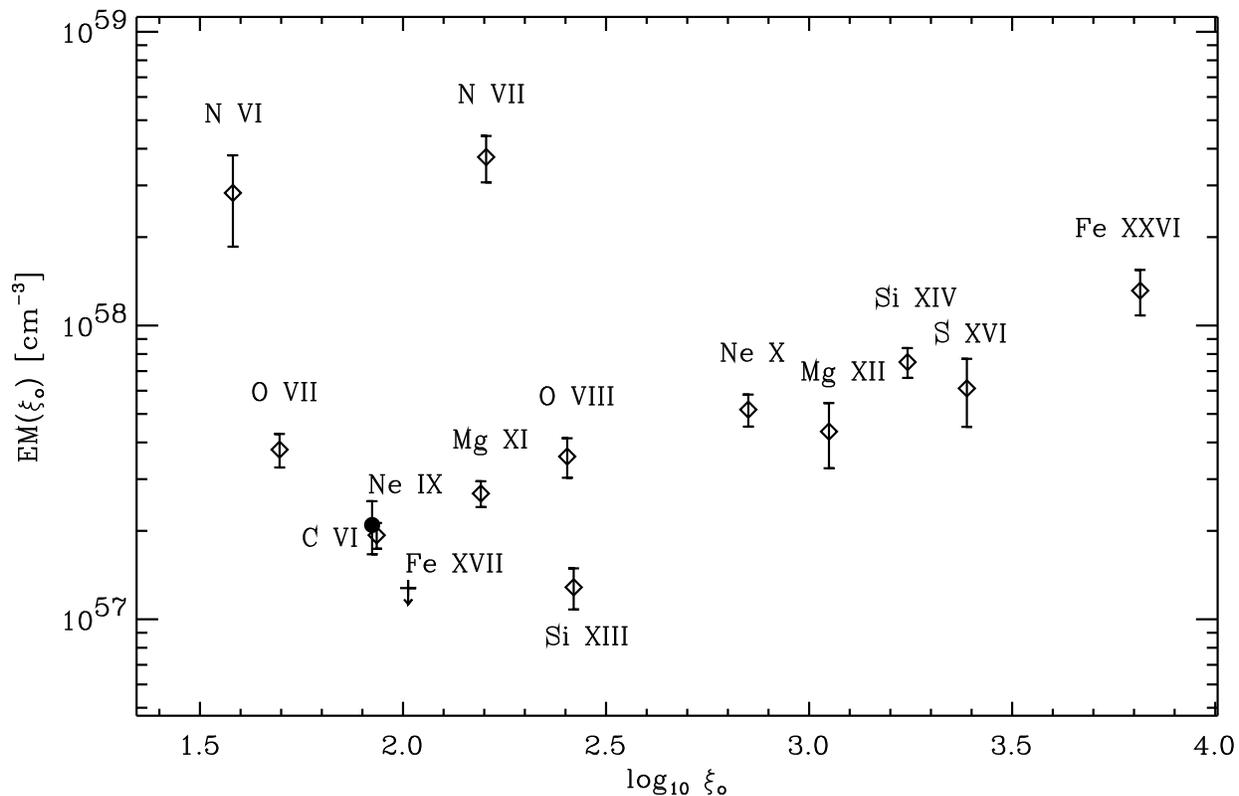}
\caption{Estimated emission measure (EM) at a single ionization parameter of formation $\xi_o$,
derived from the HETGS line fluxes (diamonds) and from the \ion{C}{6} flux measured with
RGS \cite[filled circle]{herx1me}. Both \ion{N}{7} and \ion{N}{6} indicate
an excess nitrogen abundance. 
This rough estimate of the EM has the virtue of
being insensitive to thermal instabilities. We assume a
charge state fraction $f_{i+1}=0.25$ for the recombining ions.
\label{fig:em}}
\end{figure}

\begin{figure}
\epsscale{.7}
\plotone{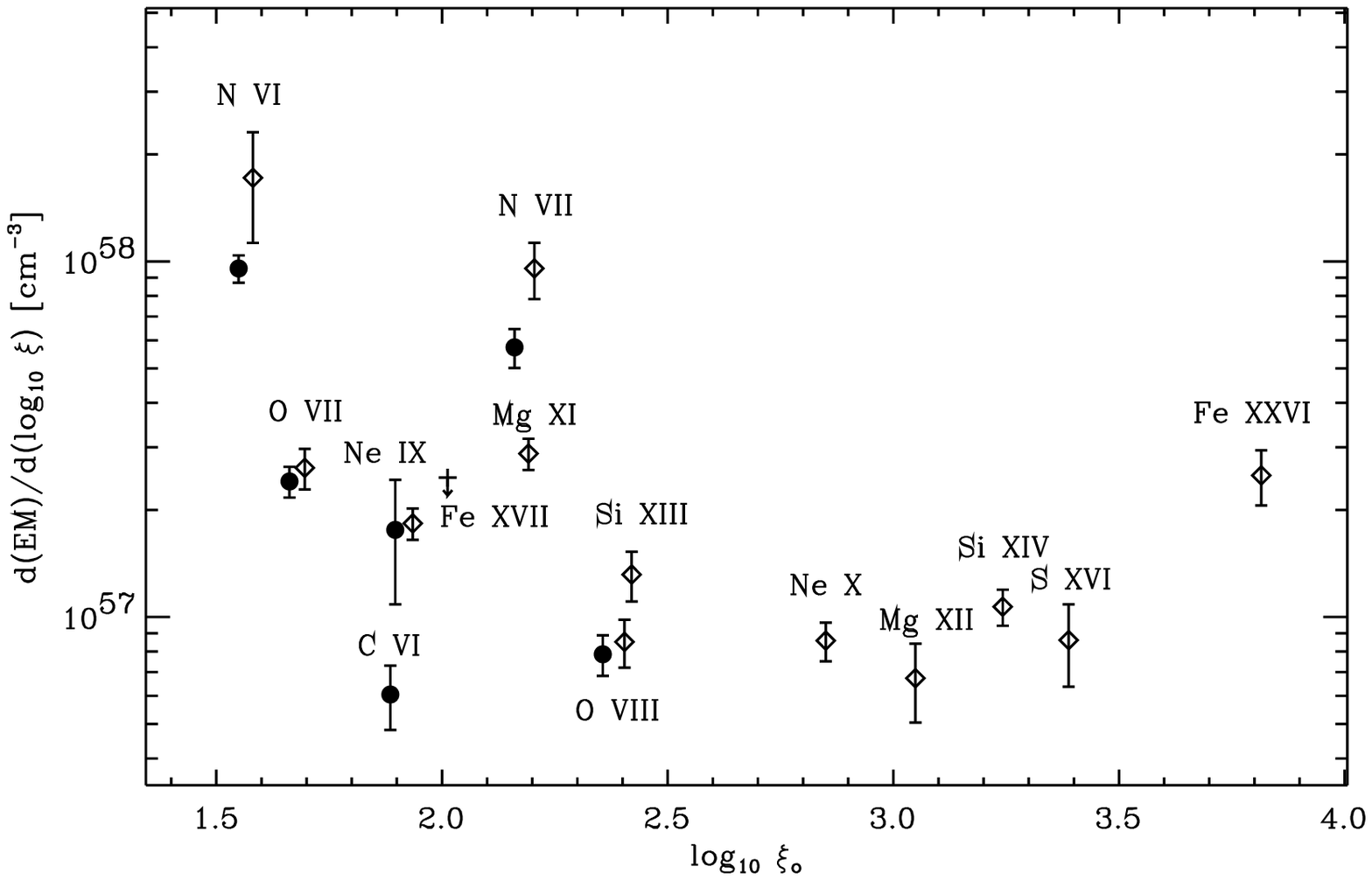}
\caption{Differential emission measure (DEM) vs. ionization parameter.
The line fluxes measured with HETGS (diamonds) are
complemented with the fluxes from \xmm RGS \cite[filled circles]{herx1me}.
This is more accurate than Fig. \ref{fig:em}
because the DEM is calculated in a grid spanning the full $\xi$ range.
However, it is also sensitive to the thermal and ionization balance solution from
XSTAR \cite[]{xstar}, as well as the DEM slope.
The over-abundance of N and depletion of C is observed. 
The predicted thermal instability regime occurs at $1.9 < \log_{10} \xi < 2.4$.
The excess DEM for the He-like ions in this regime suggests
that the plasma is more stable than predicted.
\label{fig:emcorr}}
\end{figure}

\begin{figure}
\epsscale{.7}
\plotone{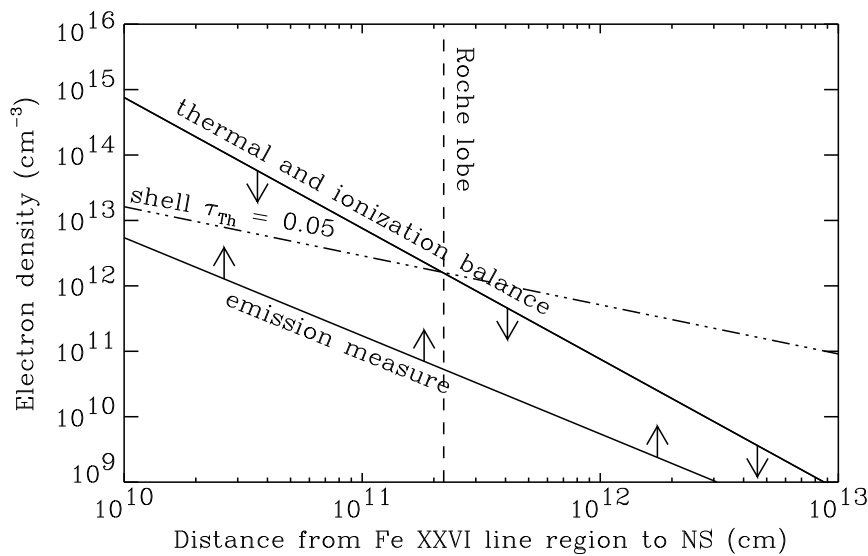}
\caption{Density limits derived for the \ion{Fe}{26} emission region, 
versus its maximum distance ($r$) to the neutron star.
The Thomson depth is plotted for a spherical shell with $r_{\rm min} = r/3$ and
$r_{\rm max} = r$.
\label{fig:fe26}}
\end{figure}

\begin{figure}
\epsscale{.7}
\plotone{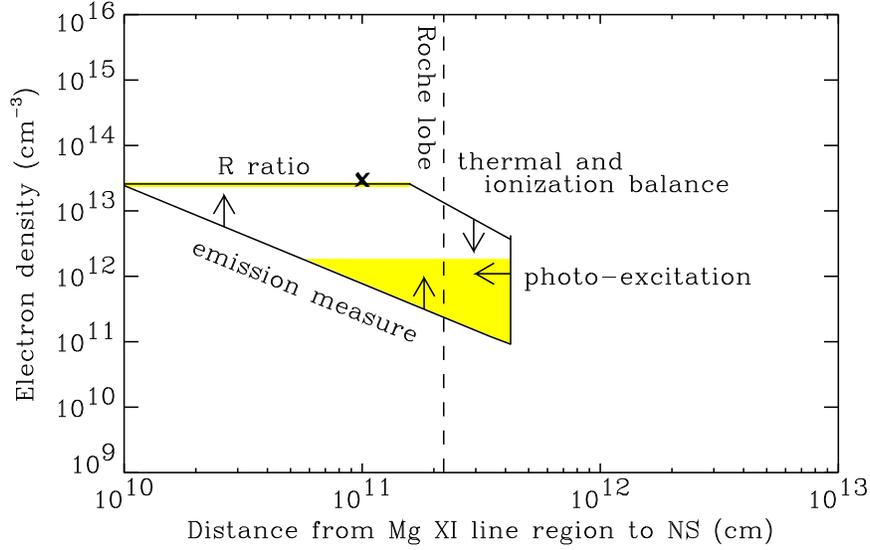}
\caption{Limits on the density of the \ion{Mg}{11} emission region, versus its maximum
distance to the neutron star. The $R$ ratio strongly limits the plasma to be close
to the density and photoexcitation boundary. 
The irradiated disk atmosphere model prediction is shown by an {\bf X},
and the thermally stable density regimes in that model are shaded in yellow. 
\label{fig:mgxi}}
\end{figure}

\begin{figure}
\epsscale{.7}
\plotone{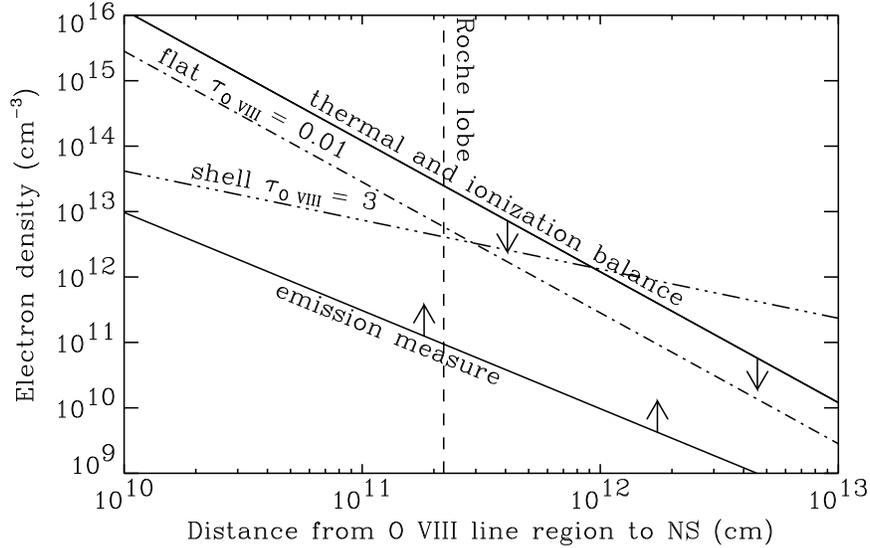}
\caption{Limits on the density of the \ion{O}{8} emission region, versus its maximum
distance to the neutron star. The velocity broadening suggests that the emission region should
be close to the Roche Lobe of the neutron star. The \ion{O}{8} edge optical depths for 
two geometries are shown: for a filled spherical shell with  $r_{\rm min} = r/3$ and
$r_{\rm max} = r$ ("shell"), and for a face-on flattened disk geometry ("flat").
\label{fig:oviii}}
\end{figure}

\begin{figure}
\epsscale{.6}
\plotone{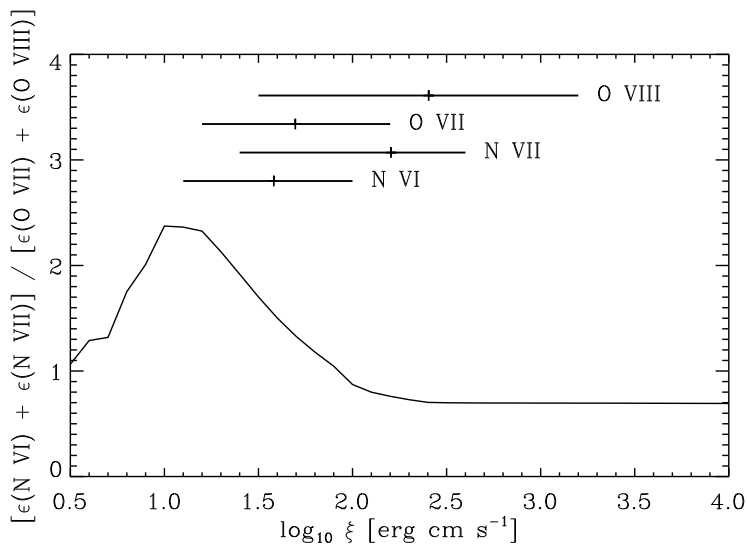}
\caption{Line emissivity ratio ($E$) built from the \ion{N}{6}, \ion{N}{7}, \ion{O}{7}, 
and \ion{O}{8} line emissivities obtained from the XSTAR and 
HULLAC models, as a function of ionization parameter $\log_{10} \xi$. 
The observed line ratio will differ from this emissivity
ratio by a factor of $A_{N}/A_{O}$, the ratio of elemental abundances.
At the top of the 
figure, we show the $\log_{10} \xi$
range at which the line emissivities are $> 20$\% of the
peak emissivity, with the $\log_{10} \xi_o$ of formation (peak emissivity) marked.  
\label{fig:emis}}
\end{figure}

\begin{figure}
\epsscale{0.4}
\plotone{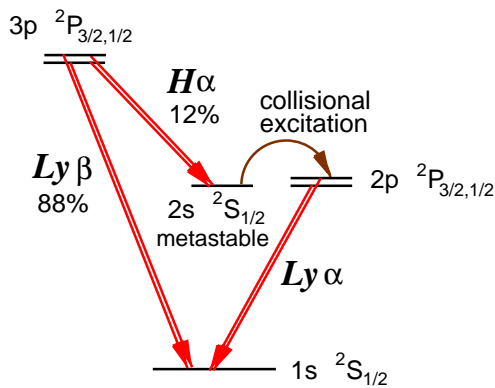}
\caption{Abridged energy level diagram for a H-like ion, showing 
the decay channels of the $3p\ ^2P_{3/2,1/2}$ levels. These channels show how
a Ly$\beta$ photon gets scattered or converted to other photons.
Once a Ly$\beta$ photon is absorbed, there is an 88\% probability 
for an electron at either of the $3p\ ^2P_{3/2,1/2}$ levels to re-emit a Ly$\beta$ 
photon, resulting in a resonant scattering event. 
Alternatively, there is a 12\% probability
for the same initial states to decay by emission of an H$\alpha$ photon.
Once an H$\alpha$ is emitted, the metastable $2s\ ^2S_{1/2}$ level decays
to the ground state via two photons (not shown), unless the electron density is 
above a critical threshold $n_{\rm crit}$. For $n_e > n_{\rm crit}$, 
the $2s\ ^2S_{1/2}$ level gets excited to
$2p\ ^2P_{3/2}$, and a Ly$\alpha$ photon is emitted.
The photons shown in the figure are all unresolved doublets. \label{fig:hlike}}
\end{figure}

\begin{figure}
\epsscale{.5}
\plotone{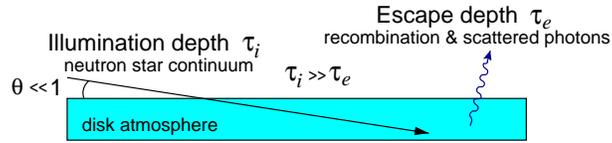}
\caption{In an accretion disk atmosphere which is illuminated with $\theta \ll 1$, 
two optical depths are relevant. In the Her X-1 spectrum, recombination emission dominates
over resonance scattering, implying that $\tau_i \gtrsim 100$ for the He-like 
ion $r$ lines. 
The mean number of scatterings and any line destruction are sensitive to $\tau_e$.
The X-ray Bowen fluorescence effect is excluded because $\langle N \rangle \lesssim 69$
and thus $\tau_e \lesssim 99$ for 
\ion{O}{8} Ly$\alpha_1$. The measured $\tau$ are beginning to show 
that the emission region is elongated.
\label{fig:depth}}
\end{figure}

In Figs. \ref{fig:fe}--\ref{fig:megon} we over-plot the synthetic spectrum
of an illuminated accretion disk atmosphere and corona model from \cite{diskmodel}.
We compare the data to this disk atmosphere model.
The model assumes that the neutron star continuum
energizes the vertically stratified structure of
a hydrostatic accretion disk in thermal and ionization balance.
We used two adjustable parameters to fit the spectrum: a normalization factor of $N = 0.11$,
and the disk radius range $10^{10.9} < r < 10^{11}$~cm (the outer annulus in the model).
Both parameters affect the normalization and do not influence the spectral shape.
We did not include any Doppler velocity broadening effects. The ionizing continuum
in the model consists of a bremsstrahlung with $T = 8$~keV and a luminosity 
$L_x = 10^{38.3}~ (D/6.6~{\rm kpc})^2 ~ N$~\ergps. Given
the approximate linear dependence of line flux on continuum flux
\cite[]{diskmodel}, the effective ionizing luminosity in the model
fit was $L_x = 2.3 \times 10^{37}$~\ergps. This is nearly the
unabsorbed luminosity during main-on state of Her X-1 in the 0.1--200~keV
band, $L=3.8 \times 10^{37}$~ergs$^{-1}$ \cite[]{beppo_cyc}.
The model normalization can be interpreted as simply a geometrical parameter which 
indicates that the observed emission is produced by an outer section in the 
accretion disk atmosphere and corona
with area $A \sim 3 \times 10^{21}$~cm$^2$.
That $A$ is a small fraction of the total disk area
is in agreement with observing an outer disk
rim, as suggested by the small velocity broadening of the emission lines (\S \ref{sec:broad}).
The neutron star continuum observed during the main-on powers
the recombination emission observed during the low state.
The precession of the accretion disk allows us to isolate
the reprocessed emission from the direct emission from the neutron star.

The relative strengths of the spectral lines, which are fixed in
the model, are a direct measure of the structure of the accretion disk atmosphere
and corona. In general, they constrain the temperature, ionization,
and density distributions. 
In the context of our model, the relative line strengths are indicators
of the vertical ionization structure of the disk atmosphere and corona, since
the ionization level and temperature are stratified with vertical depth.
In Table \ref{tab:lines}, we compare the relative strengths of lines with different 
ionization parameters (except for Fe K$\alpha$ fluorescence at 1.94~\AA, which
is fit separately, because it is not yet calculated in the disk model). We note that the
relative strengths of the Ly$\alpha$ lines of \ion{S}{16}, \ion{Si}{14}, \ion{Mg}{12}, \ion{Ne}{10},
and \ion{O}{8} lines were fit with $\sim 50$\% or better accuracy, with
{\it virtually zero degrees of freedom}, because $N$ and the radius 
range, the two free parameters in the model, only affect the overall flux normalization. 
Since \ion{O}{8} and \ion{S}{16} are produced at 
ionization parameters that are one order of magnitude apart, 
it is a notable success for the model
to be able to reproduce their flux ratios.
In the disk model context, this indicates that the emission
measure distribution of the upper half of the atmosphere agrees well with the data.
The incidence angle on the disk atmosphere is 
self-consistently determined by the model \cite[]{diskmodel}.

The \ion{Fe}{26} Ly$\alpha$ line probes the hottest and most extended region 
of the disk corona observed with HETGS (see \S \ref{sec:volden} and Fig. \ref{fig:fe}). 
The XSTAR photoionization equilibrium plasma model \cite[]{xstar}, coupled with
the recombination rates obtained with the HULLAC atomic code \cite[]{hullac}, show that 
the electron temperature of the plasma producing the \ion{Fe}{26} line is 
$T_e \sim 6.7 \times 10^6$~K, with ionization parameter $\log \xi \sim 3.8$. 
Similar parameters are obtained with the disk atmosphere model.
The observed \ion{Fe}{26} Ly$\alpha$ line flux is 3.9 times larger 
than the disk model's.
Judging from the disk model, the 
\ion{Fe}{26} H$\alpha$ and H$\beta$ recombination lines at 9.68~\AA \ and 9.53~\AA \
(see Fig. \ref{fig:megheg} and Table \ref{tab:lines}) are 10 times weaker than expected.
The flux ratio of H$\alpha$ and Ly$\alpha$ lines
is not expected to depend strongly on temperature
in the pure photoionization case. 
The \ion{Fe}{26} Ly$\alpha$ line was also detected with \xmm \cite[]{zane}.
An emission line of \ion{Fe}{25} at 1.85~\AA \ appears 
in the model, as well as a \ion{Fe}{24} complex with lines at
11.18~\AA, 11.03~\AA, and 10.62~\AA \ (these lines appear in Fig. \ref{fig:megheg}). 
None of these relatively weak lines are detected. In short, 
the upper coronal structure appears to be more ionized and to have a larger emission measure
in Her X-1 than in the disk atmosphere and corona model.

We have shown that several key characteristics of the X-ray recombination spectrum
can be explained by an outer section of the accretion disk which is being 
illuminated by the neutron star.  This disk atmosphere model describes much of the Her 
X-1 X-ray line spectrum from first principles.
However, there are discrepancies between the model 
and the data.  One notable discrepancy is \ion{N}{7} and \ion{N}{6}, which are under-predicted by
a factor of $\sim 10$. This is consistent
with the \S \ref{sec:noratio} results.
Also, the He-like ion line fluxes are systematically underpredicted by 
the model, as we discuss below.

\section{Thermal stability of the photoionized plasma and 
the H-like/He-like ion line ratios}
\label{sec:stab}

A thermal instability has been predicted to be present in plasmas which are photoionized by an 
X-ray continuum. The ``type~I'' instability is produced by
an imbalance between bremsstrahlung cooling and Compton heating, just below the Compton temperature \cite[]{kmt}.
The ``type~II'' instability is produced by the imbalance between 
recombination cooling and photoionization heating \cite[]{kallman1}.
The photoionized disk atmosphere models predict a type~II instability 
regime at $6 \times 10^4 < T < 5 \times 10^5$~K \cite[]{diskmodel}.
X-ray recombination emission is suppressed in the latter regime.
The recombination line fluxes are sensitive to the ion fraction $f_{i+1} (\xi)$.
The instability reduces the interval $[\log \xi_o-\Delta(\log\xi), ~\log \xi_o+\Delta(\log\xi)]$
at which $f_{i+1} \gtrsim 0.1$.
For this reason, the disk atmosphere model predicts 
very weak \ion{Mg}{11} line fluxes relative to \ion{Mg}{12}, 
contrary to observations (as shown in Fig. \ref{fig:megheg}). 
The DEM analysis indicates that
the \ion{Mg}{11} flux is above the expectation (Fig. \ref{fig:emcorr}). 
The ions with $\xi_o$ near the instability regime will have their line power reduced
and their derived DEM increased.

The He-like ion lines of \ion{Si}{13}, \ion{Mg}{11}, \ion{Ne}{9}, 
\ion{O}{7}, and \ion{N}{6} are underpredicted by the atmosphere model.
Also, the DEMs derived from the same He-like ions are systematically larger than
the DEMs from the H-like ions.
This may imply that the plasma is more thermally stable than the models predict,
or it may require a larger EM of $T \sim 10^5$~K plasma.
The He-like ion lines are especially sensitive to the thermal instability
because of a key physical difference: the H-like ions form by the 
recombination of fully stripped atoms, 
so their line emissivities are less sensitive to $\xi$ than those of
He-like ions. The 
$f_{i+1} (\xi)$ for the He-like lines are single-peaked functions, while 
the $f_{i+1} (\xi)$ for the H-like lines are asymptotic curves with 
$f_{i+1} \to 1$ as $\xi \to \infty$, and $f_{i+1} \to 0$ as $\xi \to 0$. 
This causes the line emissivities
for the He-like ions to have narrow peaks and small $\Delta(\log\xi)$, while
in contrast, the line emissivities 
of H-like ions have broad asymmetric peaks with 
large $\Delta(\log\xi)$, with tails that extend to large $\xi$.
This is why, in spite of a possible overlap of the $\xi$ of formation between the
He-like and H-like ions, the H-like ion line fluxes will still dominate
over the He-like ion lines at high $\xi$.
Therefore we expect the He-like ion lines to be more sensitive indicators
of the ionization distribution and the instability regime than the H-like ion lines.
The changing trends of the EM in Figure \ref{fig:em}
compared to the DEM in Figure \ref{fig:emcorr}, suggest that
the thermal instability regime is larger in the models than in
the plasma, but this is difficult to ascertain without a systematic
study of the plasma models.

The $n_e$ measurement from \ion{Mg}{11} He$\alpha$ (\S \ref{sec:volden}) also tests 
stability of the photoionized plasma, since
the disk atmosphere model predicts a regime of forbidden density. 
The $n_e$ regions allowed by both the spectral diagnostics 
and disk atmosphere models are shown in Figure \ref{fig:mgxi}.
The density of formation for \ion{Mg}{11} is $n_e = 3 \times 10^{13}$\pcmcu
according to the disk atmosphere model.
The forbidden densities are just below that, at
$2 \times 10^{12} < n_e < 3 \times 10^{13}$\pcmcu.
The measured $n_e = (2 \pm 1) \times 10^{13}$\pcmcu
is located in the high density end of the stable regime.
The thermal instability calculation
in the disk model is in agreement with the observations in this case.

The X-ray spectroscopic data is now sufficiently detailed that it
is possible to start testing the regimes of thermal instability
in the photoionized plasma.  
On the one hand, a DEM analysis based on XSTAR plasma models and the HULLAC recombination rates
suggests that the the plasma is more thermally stable than was originally predicted.
On the other hand, the density constraints from the $R$ ratio
of \ion{Si}{13} indicate that the density is just at the edge of the stability
region, as derived from the disk atmosphere model. Further, the same disk atmosphere
model predicts the \ion{Mg}{11} He$\alpha$ flux to be suppressed due to the
instability, but no such suppression is observed.

\section{X-ray Spectral Comparison with other LMXBs}
\label{sec:adc}

We compare the spectral properties of the Her X-1 low state with 
two low-mass X-ray binaries: 4U~1822-37 and 2S~0921-63.
Both 4U~1822-37 and 2S~0921-63 are classified as ADC sources.
These LMXBs exhibit X-ray emission lines with unresolved velocity broadening.
The limits on the velocity widths (FWHM) of the recombination lines measured with Chandra 
HETGS are $\Delta v < 1500$\kmps for 2S~0921-63 \cite[]{0921},
$\Delta v < 300$\kmps for 4U~1822-37 \cite[]{4u1822}, and
$\Delta v < 260$\kmps for Her X-1 
(from \ion{O}{7} in Table \ref{tab:lines}).
Her X-1 and 4U~1822-37 exhibit RRCs that indicate
that photoionization is the dominant ionizing mechanism in the plasma.  
In 2S~0921-63, the bright \ion{O}{7} intercombination line 
(giving the value of $G \sim 4$ for that diagnostic line ratio)
is indicative of a photoionized plasma \cite[]{0921}.  
The non-detection of RRCs in
2S~0921-63 is likely due to the low signal-to-noise ratio of the emission features.
The weak or undetected Fe L-shell lines in these LMXB spectra 
are yet another signature of photoionization.
It is therefore clear that the lines in these LMXBs are produced by 
recombination following photoionization. 

When looked at more carefully, a comparison of the spectra of 4U~1822-37, 2S~0921-63, 
and Her X-1 reveals a nuanced picture. 
First, the continuum level varies significantly from source to source:
4U~1822-37 has $\sim 10$ times larger continuum flux in the 0.5--7~keV band
than 2S~0921-63 or Her X-1.  The continuum in 4U~1822-37 is harder than in the other two sources.
Second, the soft X-ray line spectra in 4U~1822-37 and Her X-1 
appear to be quite similar, i.e. the fluxes of the \ion{Si}{14}, \ion{Mg}{12}, 
\ion{Mg}{11}, \ion{Ne}{10}, \ion{Ne}{9}, \ion{O}{8}, and \ion{O}{7}
lines are roughly within tens of percent of each other, albeit the exact
line ratios and density diagnostics differ.
On the other hand, 2S~0921-63 shows weaker line fluxes than 4U~1822-37 and Her X-1,
perhaps because the eclipse is blocking the source of the line emission. 
Third, the line ratios between Fe emission lines from both high and low 
ionization species are different. Her X-1 exhibits a much more intense 
Fe K$\alpha$ line from 
\ion{Fe}{1}--\ion{Fe}{13} at 1.94~\AA \ than 2S~0921-63, but a nearly equal flux of 
the \ion{Fe}{26} line at 1.78~\AA.  2S~0921-63 exhibits both
\ion{Fe}{25} and \ion{Fe}{26} lines with nearly equal fluxes, and
a Fe K$\alpha$ fluorescence line with $1/4$ of the flux of the \ion{Fe}{25}/\ion{Fe}{26} 
lines \cite[]{0921}. 
It is conceivable that the fluorescing material was occulted by the eclipse of 2S~0921-63, while
the hotter, more extended emission was not. 
4U~1822-37 also showed Fe K$\alpha$ and \ion{Fe}{26} lines, with Fe K$\alpha$ 
being brighter \cite[]{4u1822},
but it did not show the large flux contrast between those lines observed in Her X-1.
Fourth, the \ion{N}{7} Ly$\alpha$ and \ion{N}{6} 
He$\alpha$ lines are most intense in Her X-1 (see Fig. \ref{fig:megon}). 
So far, other LMXBs and ADC have not shown such strong N lines, albeit the measurement
is rendered difficult for most cases due to the large absorbing hydrogen 
columns \nH that are common in these systems.
In \S \ref{sec:em} and \S \ref{sec:cno}, we quantified this abundance
anomaly, which was also observed with the \xmm RGS \cite[]{herx1me}. 

We now compare X-ray flux variability with orbital phase.
While our observations of Her X-1 are restricted to mid-orbit at $0.33 < \phi < 0.67$,
the 2S~0921-63 observations were restricted to mid-eclipse,
and the 4U~1822-37 observations spanned full orbits. 
The emission during the 2S~0921-63 eclipse originates in
either the outer disk rim or in an extended circumsource region (such
as a corona) with an upper size limit of $7 \times 10^{10}$~cm \cite[]{0921}. 
The emission line region in 4U~1822-37 was identified to be the
illuminated disk bulge, because the line fluxes were observed to peak
at $\phi \sim 0.25$ \cite[]{4u1822}.
Flux variations were not observed in Her X-1, but with the
limited phase coverage we can ascertain that it does not show
the same sort of variation than 4U~1822-37.
Judging from the spectral variations as a function of orbital phase,
the azimuthal location of the line emission regions in
Her X-1, 2S~0921-63, and 4U~1822-37 appears to be distinct 
amongst these sources.

Her X-1, 4U~1822-37 and 2S~0921-63 are sources that 
are observed at high-inclination angles ($85$\degr$ \lesssim i < 90$\degr),
which causes the accretion disk to block the direct X-ray emission from the accreting object. 
ADCs have a higher optical-to-X-ray flux ratio than other LMXBs,
which is a well-known indicator that the accreting object is not observed directly 
\cite[and references therein]{whiter}. 
This, in turn, causes the EWs of the X-ray emission lines in ADCs to be much
larger than in other LMXBs, because only $\sim 10^{-2}$--$10^{-3}$ of 
the available continuum energy is reprocessed into recombination lines. 
This scenario is consistent with the model presented in \S \ref{sec:modcomp}, where only a fraction
of the accretion disk area can account for the line emission.
Incidentally, the only other ADC for which a high-resolution spectrum was obtained 
is AC~211 (in the M15 cluster), but it did not show emission lines 
\cite[]{white}, probably due to contamination from a bright and nearby LMXB burster. 

In sum, high-inclination LMXBs exhibit recombination-dominated line spectra and unresolved line broadening.
The illuminated disk bulge seen in 4U~1822-37, however, is not observed in Her X-1 with the current data,
and only the extended emission during eclipse was observed in 2S~0921-63.
A definitive comparison of the accretion flow geometry of these sources will require
new spectral data with good orbital phase coverage for Her X-1 and 2S~0921-63. 

\section{Conclusions}
\label{sec:concl}
\begin{enumerate}
\item We detect more than two dozen spectral features in the X-ray spectrum of the Hercules X-1 low state
with the \chandra HETGS. 
Most of the observed features correspond to the 
recombination signatures expected from a centrally illuminated 
accretion disk atmosphere and corona located in the outer accretion disk, 
given the following evidence:
\begin{itemize}
\item Using photoexcitation diagnostics with the He-like ion lines, 
coupled with the observed ultraviolet continuum fluxes by \cite{fuse}, we ruled out the possibility 
that the {\it recombination} emission lines originate
on the illuminated companion star, HZ~Her. 
\item The spectral model for a centrally illuminated accretion disk
atmosphere and corona fit most of the observed fluxes and ratios of the X-ray recombination lines.
The model requires the low state emission to be energized by the continuum observed in the high state.
The caveat is that the He-like lines are underpredicted by the model,
and the observed line broadening is smaller than
expected for a plasma corotating with the accretion disk. 
The small Doppler widths 
may be due to vignetting effects that allow only the far edge of the disk to be visible, 
as would occur for an edge-on disk with a flared or warped geometry.
In support of this, the model fits indicate that only the 
outer rim of the disk ($8 \times 10^{10} < r < 10^{11}$~cm) is observed, in contrast 
with LMXBs such as EXO~0748-67, where much of the inner disk emits
observable X-ray lines \cite[]{exo}.
\item The measured $R$ line ratio of \ion{Mg}{11} allows us to precisely determine a 
density of $n_e = (2 \pm 1) \times 10^{13}$\pcmcu, which agrees with the disk atmosphere 
model predictions ($n_e = 3 \times 10^{13}$\pcmcu). 
This corresponds to the low UV field limit, which is the most likely scenario for \ion{Mg}{11}.  
Using UV observations \cite[]{fuse}, we deduce that
the disk may be far enough from HZ~Her for the \ion{Mg}{11} triplet to be unaffected by
photoexcitation.
The presence of UV emission 
and high-$n_e$ necessitates the re-calculation of $R$ ratios in $(n_e, F_{\rm UV})$ space.
\item The X-ray spectrum probes radically different length scales in
the atmosphere and corona. For example, an 
emission measure analysis shows that the volume where \ion{Fe}{26} is emitted
must be $\sim 10^4$ times larger than that for \ion{N}{6}.
Compare this to the 
cool model atmosphere emitting \ion{N}{7} at $T \sim 32500$~K, with $10^7$~cm thickness, 
and $3 \times 10^{13}$\pcmcu density, which coexists with a $10^{10}$~cm thick hot
corona emitting \ion{Fe}{26}, with $10^{11}$--$10^{12}$\pcmcu density and $T \sim 10^6$~K 
(with intermediate layers of ionization in between). 
\item The RRC/RR ratios and the measured $T = 81000 \pm 35000$~K from
\ion{Ne}{9} RRC are generally in agreement with the disk atmosphere model. Single-$T$
models do not reproduce well the 
observed RRC/RR flux ratios of H-like ions, while the disk model produces
improved agreement. This indicates the presence of multiple temperature components, since
any high-$T$ component of the RRCs is lost in the continuum.
\item The line depths derived from the 
spectrum via the Ly$\alpha$/Ly$\beta$ and $G$ ratios
are consistent with the ionic column densities and flattened geometry of
the model disk atmosphere, where the first ratio is most sensitive to
escape depth (vertical depth) and the second is determined
by the grazing incidence depth.  To obtain consistency, we use a turbulent velocity width 
of $\Delta v \sim 150$\kmps.  We rely on the relationship between $\tau$ and the mean number
of scatters $\langle N \rangle$ calculated for a range of Voigt parameters by \cite{hummer}.
\end{itemize}
\item We find a prominent Fe K$\alpha_1,\alpha_2$, K$\beta$ complex indicative of a
neutral plasma.  We cannot ascertain whether the
Fe K fluorescence is produced in the disk or
on HZ~Her, or both. The Fe K$\alpha$ line flux from HETGS does not vary with orbital phase at 
$0.3 \lesssim \phi \lesssim 0.7$, but \cite{zane} did observe a clear modulation 
with $\phi$ with \xmms.
\item The lines in the X-ray spectrum have unresolved velocity broadening ($\Delta v < 260$\kmps),
similar to what has been observed for the lines in ADC sources.
This is evidence that the accretion disk 
is edge-on with respect to our line of sight.
Previously observed spectral changes with the \xmm RGS showed a transition between
broad lines in the main-on state and narrow lines in the short-on and low states, which
suggested emission from the inner disk during main-on and from the outer disk during short-on and low state
\cite[]{herx1me}.  The well-known 35~d flux modulation has long been attributed
to the precession of an edge-on disk.
\item By measuring the mean number of line scatterings, we rule out 
the presence of the X-ray Bowen
fluorescence effect calculated by \cite{sakorad}.
We constrain the mean number of scatterings of the \ion{O}{8}
Ly$\alpha_1$ line to $\langle N \rangle \lesssim 69$. 
The upper limit is based on the detection of \ion{O}{8} and \ion{Ne}{10}
Ly$\beta$ lines, which must be
destroyed after a large number of scatterings. 
In contrast, the X-ray Bowen fluorescence effect would 
require $\langle N \rangle \sim 5 \times 10^3$.
Therefore, the \ion{N}{7}/\ion{O}{8} Ly$\alpha$ line ratio is unaffected by optical
depth effects. 
\item The signature of CNO abundances observed with \xmm \cite[]{herx1me} is 
confirmed.  Strong \ion{N}{6} He$\alpha$, \ion{N}{7} Ly$\alpha$, and 
\ion{N}{7} RRC lines are present in the HETGS and in the \xmm RGS spectra.
The N abundance is verified to be higher than Solar. 
The simplest EM analysis, the DEM analysis, and the 
(\ion{N}{7} + \ion{N}{6})/(\ion{O}{7} + \ion{O}{8}) line ratio yield
$4 \lesssim ({\rm N/O})/({\rm N/O})_\sun \lesssim 9$. The uncertainty is
dominated by systematics in the analysis rather than by the effective area calibration.
The depletion of carbon derived from the RGS data is
also evidenced in our DEM analysis.
This is supported by measurements of the N/C ratio made by 
comparing the observed UV spectrum to that from an illuminated accretion
disk model \cite[]{raymond}.
The relative O abundance is sensitive to the Solar abundance values, being
under-abundant with 
the \cite{wilms} values, and consistent with Solar with the
\cite{coabun} values.
\item Given the problems with fitting the He-like ion line fluxes with either the disk model or the DEM, we deem
that either a) the thermal instability regime is over-predicted; or 2)
the disk atmosphere model requires more emission measure at $T \sim 10^4$-$10^5$~K. 
The He-like ion lines are more 
sensitive probes of the thermal equilibrium of the
plasma than the H-like ion lines.
It is possible that the thermal instability has been over-estimated in
the models, therefore cutting the He-like ion fluxes disproportionately.
However, the density diagnostics seem to be consistent with the 
calculated stable density values.
\item There are line transfer effects which remain to be accounted for in the models, such as
the competition between line emission by photoexcitation and line destruction,
and the effects of repeated photoionizations and recombinations produced by lines and RRC.
However, the Her X-1 spectrum indicates that these are second-order effects,
compared to the pure recombination emission.
\item The atmospheric and coronal components unveiled by the X-ray line spectra
are more extended ($10^{10}-10^{11}$~cm), cooler, less luminous,
and have smaller vertical Thomson depth ($\tau_T \sim 0.01$) than the compact 
($10^9$~cm) ADC component with $\tau_T \gtrsim 1$, which resides close to the compact object.
Indeed, {\it both} compact and extended ADC components are evident in the eclipse 
ingress light curves observed in Hercules X-1 \cite[]{eclipses_choi}.
Therefore the plasma observed in high resolution spectra is not part of what
is commonly referred to as an ADC (i.e. a fully ionized spherical distribution of plasma
that is Thomson scattering the neutron star X-rays), 
but instead may be a continuation of it at the Roche lobe scale. 
The extended ADC component observed in eclipse coincides with our size measurements
for the X-ray line region. 
\end{enumerate}

\acknowledgments

\clearpage



\clearpage

\clearpage

\clearpage

\clearpage

\clearpage

\clearpage

\clearpage

\clearpage

\clearpage

\begin{deluxetable}{lrrrrl}
\tabletypesize{\scriptsize}
\tablecaption{X-ray Emission Features and Disk Atmosphere Model Predictions
\label{tab:lines}}
\tablewidth{0pt}
\tablehead {
\colhead {Line(s)} & $\lambda$ [\AA]\tablenotemark{a} & $(\Delta v)_\sigma$ [km/s] & Observed Flux\tablenotemark{b} & Model Flux\tablenotemark{b}~($\lambda$)  }
\startdata
N VI $i$       &     29.08   &  \nodata          & $      64 \pm     22  $  & 1.9 \\
N VII Ly$\alpha$&    24.78   & \nodata      & $      25 \pm     4.5  $  & 2.0 \\
O VII $f$      &     22.10   & \nodata           & $  1.2 \pm 1.2 ~( < 3.9 ) $ & 0.36  \\
O VII $i$      &     $21.800 \pm 0.006$  & $< 110 $		& $      40.5 \pm     5.5 $ & 9.2   \\
O VII $r$      &     21.60   & \nodata           & $       3.1 \pm       1.6 $  &  2.5  \\
O VIII Ly$\alpha$   &     $18.970 \pm 0.006$   & $ < 270 $    & $      13 \pm 2 $ & 9.0  \\
N VII RRC      &     18.59  & \nodata           & $  18 \pm 5  $  & 0.63 \\
Fe XVII        &     17.05, 17.10  & \nodata           & $      < 3.8 $ & 0.56  \\
O VII RRC      &     16.78  & \nodata           & $2.5 \pm 1.0$ & 4.2 \\
O VIII Ly$\beta$ &     16.01   & \nodata   & $0.85 \pm 0.30 $  & 2.0  \\
O VIII RRC      &     14.23 & \nodata           & $5.8 \pm 1.2$   & 2.4 \\
Ne IX $f$      &     13.70   & \nodata           & $  0.17 \pm 0.17 ~( < 0.96) $  & 1.4  \\
Ne IX $i$      &     $13.553 \pm 0.003$   & $< 160$           & $       7.3 \pm     0.75 $ & 0.71 \\
Ne IX $r$      &     13.45   & \nodata           & $       0.87 \pm       0.33 $  & 0.66 \\
Ne X Ly$\alpha$     &     $12.135 \pm 0.003$  &  $< 240$     & $       3.2 \pm       0.4 $ & 1.8  \\
Ne IX RRC      &      10.37  &    \nodata        & $1.82 \pm 0.36$   & 0.80 \\
Ne X Ly$\beta$      &     10.24   & \nodata      & $       0.59 \pm       0.16 $  & 0.67   \\
Fe XXVI H$\alpha$      &   9.53, 9.67 &  \nodata          & $ < 0.24 $  & 0.79 \\
Mg XI $f$      &      9.31   &  \nodata          & $       0.84 \pm       0.17 $  & 0.32  \\
Mg XI $i$      &      $9.229 \pm 0.003$   & $ < 380 $         & $       1.65 \pm       0.21 $ & 0.14   \\
Mg XI $r$      &      9.17   &    \nodata        & $       0.59 \pm       0.15 $  & 0.10 \\
Ne X RRC       &      9.10   &    \nodata        & $0.73 \pm 0.19$  & 0.32 \\
Mg XII Ly$\alpha$   &      $8.417 \pm 0.003$   & \nodata      & $  1.12 \pm  0.28 $  & 1.1  \\
Mg XII Ly$\beta$    &      7.11   & \nodata      & $       0.42 \pm       0.11 $  & 0.30   \\
Si XIII $f$    &      6.74   & \nodata           & $       1.08 \pm       0.15 $ & .38 \\
Si XIII $i$    &      6.69   & \nodata           & $   0.36 \pm 0.12 $  & .18 \\
Si XIII $r$    &      $6.647 \pm 0.003$   & \nodata           & $       0.54 \pm       0.13 $  & .11 \\
Si XIV Ly$\alpha$   &      $6.185 \pm 0.003$   & $< 340$      & $       1.44 \pm       0.23 $  & 1.8 \\
Si XIV Ly$\beta$    &      5.22   & \nodata      & $       0.61 \pm       0.26 $ & 0.35 \\
S XVI Ly$\alpha$    &      4.73   & \nodata      & $       1.19 \pm       0.31 $ & 0.88 \\
Fe II--XIII K$\alpha_2$      &      $1.940 \pm 0.003$\tablenotemark{c} & $ < 680 $    & $      10.61 \pm       1.84 $  & \nodata  \\
Fe II--XIII K$\alpha_1$      &      $1.936 \pm 0.003$\tablenotemark{c} & $ < 680 $    & $      22.00 \pm       1.94 $  & \nodata  \\
Fe XXVI Ly$\alpha$  &      1.78   & \nodata      & $       7.37 \pm       1.30 $   & 1.9  \\
Fe II--XIII K$\beta$         &      1.76   & \nodata      & $       5.10 \pm      1.20$ & \nodata     \\
\enddata

\tablecomments{The model corresponds to an accretion disk atmosphere and corona.
The statistical errors and upper limits 
are given to 66~\% confidence, except for the upper limits, which are 90~\% confidence.
Symbols: $\lambda$ = line wavelength; $(\Delta v)_\sigma$ = standard deviation of the Gaussian emission line. 
We include systematics in the measurements, except for the line fluxes. We compare the
data to the disk atmosphere and corona model. }
\tablenotetext{a}{Weak features are assigned nominal wavelengths. Otherwise, the measured
wavelengths are shown with error bars. The nominal wavelenghts are rounded up
to 0.01 m\AA. }
\tablenotetext{b}{In units of 10$^{-5}$ photons cm$^{-2}$s$^{-1}$. }
\tablenotetext{c}{This is the
relative shift of the Fe K$\alpha$ pair with wavelength and flux ratios tied.}

\end{deluxetable}

\clearpage

\begin{table}
\begin{center}
\small
\caption{Continuum fit parameters \label{tab:cont}}
\begin{tabular}{llr}
\\
\tableline\tableline
Component (Model) & Parameter (Unit) 							& Value \\
\tableline
Absorption (1) & $N_{H}$ ($10^{18}$ cm$^{-2}$) 					& 13 (fixed)	\\
Power-law (1) & Norm. at 1~keV (phot~keV$^{-1}$ cm$^{-2}$ s$^{-1}$)  	&  $(9.6 \pm 0.4) \times 10^{-4}$  \\
Power-law (1) & Photon Index									& $0.27 \pm 0.03$	   \\
Blackbody (1) & Norm. ($10^{39}$\ergps / $(10$~kpc$)^2$)			& $(3.5 \pm 0.2) \times 10^{-5}$ \\
Blackbody (1) & Temperature (keV) 								& $0.181 \pm 0.015$ \\
(1)   & $\chi^2$/DOF 										& $1292/1844 = 0.70$ \\
Absorption (2) & $N_{H}$  ($10^{18}$ cm$^{-2}$) 				& 13 (fixed)	\\
Power-law (2) & Norm. at 1~keV (phot~keV$^{-1}$ cm$^{-2}$ s$^{-1}$)  	&  $(1.34 \pm 0.01) \times 10^{-3}$  \\
Power-law (2) & Photon Index 									& $0.53 \pm 0.01$	   \\
(2) & $\chi^2$/DOF 											& $1703/1844 = 0.92$ \\
\tableline
\end{tabular}
\tablecomments{DOF = degrees of freedom. }
\end{center}
\end{table}

\begin{table}
\begin{center}
\scriptsize
\caption{Helium-like line diagnostics and photoexcitation of the $1s2s\ ^3S_1$ level 
\label{tab:helike}} 
\begin{tabular}{ccccccccc}
\\
\tableline\tableline
    & $R=f/i$  & $G=(f+i)/r$ & $\lambda_{\rm f \to i}$ & Flux $F_{\lambda_{\rm f \to i}}$  & $w_{\rm f}$  & $w_{\rm f \to i}$($r$=$10^{11}$cm) & $d$ &	$n_e$ \\
Ion & Line   & Line  &  (\AA)$^{(1)}$ & ($10^{-13}$ erg cm$^{-2}$  & $2~^3S_1 \to\ 1~^1S_0$ & $2~^3S_1 \to 2~^3P_{0,1,2}$ & Radius & Density \\
    & Ratio  & Ratio &  					 & s$^{-1}$ \AA$^{-1}$ )     & Rate ($s^{-1}$)$^{(2,5)}$ & Rate ($s^{-1}$) & (cm) & (\pcmcu)\\
\tableline
Si XIII & $3.0 \pm 1.0$  & $2.7 \pm 0.7$        	&  864 & $\sim 1.1^{(4)}$    & $3.56 \times 10^5$ & $7 \times 10^4$ & $> 8 \times 10^{10}$ & $< 5 \times 10^{12}$ \\
Mg XI  & $0.51 \pm 0.12$ & $4.2 \pm 1.2$    & 1033 & $\sim 3.1^{(4)}$    & $7.24 \times 10^4$ & $4 \times 10^5$ & $2 \times 10^{11}$ & $(2  \pm 1) \times 10^{13}$ \\
Ne IX  & $< 0.13$  & $8.6 \pm 3.4$       		& 1270 & $1.9 \pm 0.1^{(3)}$ & $1.09 \times 10^4$ & $5 \times 10^5$ & $< 6 \times 10^{11}$ & $> 1 \times 10^{13}$ \\
O VII  & $< 0.10$  & $13 \pm 7$  			& 1637 & $1.5 \pm 0.2^{(3)}$ & $1.04 \times 10^3$ & $1 \times 10^6$ & $< 3 \times 10^{12}$ & $> 8 \times 10^{11}$ \\
\tableline
\end{tabular}

\tablecomments{The statistical errors are calculated to 66\% confidence and the limits to
90\%. 
Symbols: $\lambda_{\rm f \to i}$ = wavelengths of $1s2s\ ^3S_1 \to 1s2p\ ^3P_{0,1,2}$ transitions; $w_{\rm f}$ = radiative decay rates
for $1s2p\ ^3S_1 \to 1s^2\ ^1S_0$ ;
$w_{\rm f \to i}$ = calculated photoexcitation rate
for $1s2p\ ^3S_1 \to 1s2p\ ^3P_{0,1,2}$. The energy level notation is abridged in the table header.} 
\tablerefs{(1) Porquet et al. 2001; (2) Drake 1971; (3) from the \it Hubble \rm $GHRS$ at $\phi=0.56$--$0.60$ 
by Boroson et al. 1996; (4) Boroson et al. 2001, and private communication 2003, FUSE data
near $\phi = 0.5$; (5) From HULLAC atomic code, Klapisch et al. (1977).}
\end{center}
\end{table}

\begin{deluxetable}{lccc}
\tablecaption{Ly$\alpha$ to Ly$\beta$ Line Ratios
\label{tab:lyb}}
\tablewidth{0pt}
\tablehead{
\colhead {
Ion} & 	$kT_o$  	& Ly$\alpha$/Ly$\beta$  & Ly$\alpha$/Ly$\beta$  \\
	& 	[eV]		& theory  & observed \\}
\startdata
\ion{O}{8}   & disk	 & 4.7 &  $15 \pm 6$ \\
		   & 25  & 4.9 & \\
\ion{Ne}{10} & disk	 & 3.3 & $5.4 \pm 1.6$ \\
		   & 90 	 & 4.6 & \\
\ion{Mg}{12} & disk  & 4.8 & $2.7 \pm 1.0$ \\
		   & 154 	 & 4.8 & \\
\ion{Si}{14} & disk  & 5.5 & $2.4 \pm 1.1$ \\
		   & 237 	 & 4.8 & \\
\enddata
\tablecomments{The theoretical values were calculated with HULLAC in the optically thin limit at
the temperature of peak emission, and for the disk atmosphere (with temperatures $2 < kT < 860$~eV), which includes
possible line blends with other species. $T_o$ = average temperature weighted by the emissivity of the RR line ($f_{i+1}~S_{ul}$).}
\end{deluxetable}

\begin{deluxetable}{lccc}
\tablecaption{RRC to Radiative Recombination (RR) Line Ratios 
\label{tab:rrc}}
\tablewidth{0pt}
\tablehead{
\colhead {
Ion} & 	$kT_o$  	& RRC/RR  & RRC/RR  \\ 
	& 	[eV]	& theory  & observed \\ }
\startdata
\ion{N}{6} & 2.8 	& 0.24 	& \nodata \\
	   & disk 	& 0.26 	&  \\
\ion{N}{7} & 14 	& 0.83 	& $0.72 \pm 0.25$,  $0.28 \pm 0.14^{\rm (1)}$ \\ 
	   & disk 	& 0.32 	&  \\
\ion{O}{7} & 3.5 	& 0.32	& $0.06 \pm 0.02$ \\
	   & disk 	& 0.35 	&  \\
\ion{O}{8} & 25 	& 0.87	& $0.45 \pm 0.12$ \\
	   & disk 	& 0.27 	&  \\
\ion{Ne}{9} & 6.5 	& 0.25 	& $0.22 \pm 0.05$ \\
	   & disk 	& 0.29 	&  \\
\ion{Ne}{10} & 90 	& 1.02 	& $0.23 \pm 0.07$ \\
	   & disk 	& 0.18 	&  \\
\enddata
\tablecomments{The theoretical values were calculated with HULLAC in the optically thin limit.
For RR, we use either Ly$\alpha$ or He$\alpha$, corresponding to each H-like or He-like ion, respectively. $T_o$ = average temperature weighted by the emissivity of the RR line.
The disk model temperature range is $2 < kT < 860$~eV.}
\tablerefs{(1) Jimenez-Garate et al. (2001).}
\end{deluxetable}







\end{document}